\documentclass[usenatbib]{mnras}

\usepackage{graphicx}
\usepackage{hyperref}
\usepackage{xspace}
\newcommand{\boldsymbol}[1]{\mbox{\boldmath{${#1}$}}}

\def\mdm{M_{\mathrm{DM}}}
\def\mhalo{M_{\mathrm{h}}}
\def\reff{R_{\mathrm{eff}}}

\def\ximin{\xi_{\mathrm{min}}}
\def\mtrue{M_*^{\mathrm{true}}}

\def\msalp{M_*^{\mathrm{Salp}}}

\def\aimf{\alpha_{\mathrm{IMF}}}
\def\loga0{\log{\alpha_0}}
\def\mdm{M_{\mathrm{DM},5}}
\def\sigmaee{\sigma_{e2}}

\def\pr{{\rm Pr}}
\def\hyperp{\boldsymbol{\eta}}
\def\indpar{\boldsymbol{\omega}}
\def\indpari{\boldsymbol{\omega}_i}
\def\msalpi{M_{*,i}^{\mathrm{Salp}}}
\def\aimfi{\alpha_{\mathrm{IMF},i}}
\def\mdmi{M_{\mathrm{DM},5,i}}
\def\sigmaeei{\sigma_{e2,i}}
\def\datad{\mathbf{d}}
\def\datadi{\mathbf{d}_i}

\def\Sref#1{Section~\ref{#1}\xspace}
\def\Fref#1{Fig.~\ref{#1}\xspace}
\def\Tref#1{Table~\ref{#1}\xspace}
\def\Eref#1{equation~(\ref{#1})\xspace}

\begin{document}

\title[Evolution of the stellar initial mass function of early-type galaxies]{Merger-driven evolution of the effective stellar initial mass function of massive early-type galaxies}
\author[Sonnenfeld et al.]{
Alessandro~Sonnenfeld,$^{1,2}$\thanks{E-mail:alessandro.sonnenfeld@ipmu.jp}
Carlo~Nipoti,$^{3}$
and Tommaso~Treu$^{2}$
\\
$^{1}$Kavli IPMU (WPI), UTIAS, The University of Tokyo, Kashiwa, Chiba 277-8583, Japan \\
$^{2}$Department of Physics and Astronomy, University of California, Los Angeles, 430 Portola Plaza, Los Angeles, CA 90025, USA \\
$^{3}$Department of Physics and Astronomy, Bologna University, viale Berti-Pichat 6/2, 40127 Bologna, Italy
}

\maketitle

\begin{abstract}
The stellar initial mass function (IMF) of early-type galaxies is the
combination of the IMF of the stellar population formed in-situ and
that of accreted stellar populations.  Using as an observable the
  effective IMF $\aimf$, defined as the ratio between the true stellar
  mass of a galaxy and the stellar mass inferred assuming a Salpeter
  IMF, we present a theoretical model for its evolution as a result
of dry mergers.  We use a simple dry merger evolution model, 
  based on cosmological $N$-body simulations, together with
empirically motivated prescriptions for the IMF to make predictions
for how the effective IMF of massive early-type galaxies changes from
$z=2$ to $z=0$.  We find that the IMF normalization of individual
galaxies becomes lighter with time. At fixed velocity dispersion, $\aimf$ is predicted to be constant with redshift. Current constraints on the
evolution of the IMF are in slight tension with this prediction, even though 
systematic uncertainties prevent a conclusive statement.  The correlation of $\aimf$ with stellar mass becomes
  shallower with time, while the correlation between $\aimf$ and
  velocity dispersion is mostly preserved by dry mergers.  We also
find that dry mergers can mix the dependence of the IMF on stellar
mass and velocity dispersion, making it challenging to infer,
  from $z=0$ observations of global galactic properties, what is the
  quantity that is originally coupled with the IMF.
\end{abstract}


\section{Introduction}\label{sect:intro}

Understanding the properties and the origin of the stellar
  initial mass function (IMF) is currently one of the biggest
  challenges in galaxy formation theory.  Observational constraints
on the IMF provide us with a puzzling scenario. On the one hand the
IMF appears to be remarkably self-similar across different
environments within the Milky Way \citep[see e.g.][]{BCM10, Off15}. On
the other hand, the IMF in early-type galaxies is inferred to vary
systematically as a function of mass or velocity dispersion
\citep{Tre++10,Cap++12,CvD12, Dut++12, TRN13, Spi++14, Pos++15},
although even for massive early-type galaxies we are still far from a
clear picture \citep{SLC15}.  Efforts have been put into reproducing
from a theoretical standpoint the observed IMF trends. However,
despite recent progress \citep{H+C11,Kru11,Hop12,GKH16}, we still lack a coherent description of
star formation across all environments.

One complication in comparing measurements of the IMF with models is
that present-day stellar populations are ensembles of stars formed at
different epochs in a range of environments.  For massive early-type
galaxies, a significant fraction of their present-day stellar mass is
believed to be accreted from other systems \citep[e.g.][]{vDo++10}.
If the IMF is not universal, then each accreted object will in general
have a different IMF from the preexisting population of the central
galaxy.  The IMF of a massive galaxy at $z=0$ will then be the
combination of the IMF of the stellar population formed in-situ and
that of the accreted galaxies, possibly resulting in spatial gradients
\citep{Mar++15, LaB++16}.  How does this "effective" IMF evolve in
time?  Answering this question and comparing the predictions to
observations provides a new way to test galaxy formation and star
formation models.

While the IMF is typically assumed as universal in cosmological
simulations, there are studies of galaxy evolution based on
semi-analytical models (SAMs) that allow for a non-universal IMF
\citep{Nag++05, Bek13, Cha++15, GargiuloI++15, Fon++16}. These works
explore mostly the effect of the IMF on the chemical evolution of
galaxies.  

In order to isolate the effects of a varying  IMF in the context of dry mergers, we adopt a simple model based on cosmological numerical simulations, galaxy-galaxy mergers simulations, and empirical prescriptions for the varying IMF. In practice, we use a simple
prescription for assigning the starting ($z=2$)
  IMF of an ensemble of galaxies and then evolve the stellar
population of central galaxies by merging their stellar content with
that of accreted satellites. We tune our model to match the
  correlation between IMF normalization and stellar mass observed at
  $z\sim0$ and use it to make predictions on the stellar IMF of
  massive galaxies at higher redshifts. Though very simple in its
construction, our model allows us to clearly isolate the effect of dry
mergers, which are believed to represent the main growth mechanism of
massive early-type galaxies at $z < 2$, on the evolution of the IMF.


The paper is organized as follows.  In \Sref{sect:model} we describe
our model for the IMF of $z=2$ galaxies and its evolution as a result
of dry mergers.  In \Sref{sect:obs} we present low-$z$ measurements of
the IMF used to calibrate our model.  In \Sref{sect:results} we show
our predictions on the time evolution of the IMF.  We discuss our
results in \Sref{sect:discuss}, while \Sref{sect:concl}
concludes. We assume a flat $\Lambda$CDM cosmology with $H_0=70$~km~s$^{-1}$~Mpc$^{-1}$ and $\Omega_M = 0.3$. Throughout the paper velocity dispersions are expressed in units of km~s$^{-1}$.


\section{The model}\label{sect:model}

\subsection{Parameters and notation}

Throughout our work we use the following quantities to describe the
stellar content of our galaxies. We first define a true stellar mass,
$\mtrue$. Then we introduce a Salpeter \citep{Sal55} stellar
mass, $\msalp$, defined as the stellar mass one would infer by fitting
a stellar population synthesis model based on a Salpeter IMF to
broad-band photometric data. This quantity is typically used when
observationally measuring stellar masses.  We then consider the {\em
  IMF mismatch parameter} \citep{Tre++10}
\begin{equation}\label{eq:aimf}
\aimf = \frac{\mtrue}{\msalp}.
\end{equation}
Stellar populations with a more bottom-light IMF than a Salpeter IMF
will have $\aimf<1$. A Chabrier IMF for example corresponds typically
to a value $\aimf\approx0.6$.  As defined above, $\aimf$ is a
well-defined quantity also for galaxies that do not have a homogeneous
stellar population, for example as a result of mergers.  For typical
cases $\aimf$ is mostly sensitive to the behaviour of the IMF at small
masses, which dominate the mass budget but contribute little to
  the light. Depending on the parametrization it can be a probe of
low-mass cutoff, and/or shape of the IMF. In
  extreme cases, $\aimf$ can be affected by the IMF at high masses,
  which, however, leaves an imprint not only on the mass-to-light
  ratio, but also on other phenomena, such as metal enrichment or
  supernovae rates. In this work, $\aimf$ is the only parameter used
to describe the IMF of a galaxy. Throughout our analysis we identify
the terms {\em effective IMF} and {\em IMF normalization} with this quantity.

In addition to the stellar mass of a galaxy, we track its halo
  mass $\mhalo$ and its central stellar velocity dispersion $\sigma$:
  $\mtrue$, $\msalp$, $\mhalo$ and $\sigma$ are the only quantities
  that enter our model.

\subsection{The mock sample}

We generate a sample of 10000 halos at $z=2$ with masses drawn from the
  halo mass function described by \citet{Tin++08}, using an upper cut
  off at $\log{\mhalo/M_\odot} < 13.5$.  We make this cut to focus on
the regime of galaxy or galaxy group scale halos, where most
observational constraints on the IMF are measured \citep{Gav++07},
excluding cluster of galaxy environments \citep[though IMF constraints
  are available for some brightest cluster galaxies;][]{New++13}.  We
then assign a Salpeter stellar mass to each halo using the
stellar-to-halo mass relation (SHMR) from \citet{Lea++12}, including
scatter. 
The \citet{Lea++12} SHMR is originally expressed in terms of stellar masses calculated assuming a Chabrier IMF. 
For massive red galaxies, the assumption of a Salpeter IMF in stellar population synthesis measurements produces stellar masses larger than Chabrier by $0.25$~dex, with negligible scatter \citep{Aug++10}. 
Therefore we otain an SHMR in terms of $\msalp$ by applying the transformation
\begin{equation}
\log{M_*} \rightarrow \log{M_*} + 0.25
\end{equation}
to the \citet{Lea++12} SHMR.

We make an additional cut in stellar mass keeping only
galaxies with $\log{\msalp/M_\odot} > 10.5$ and then select a random sample of
100 objects.  Finally we assign central stellar velocity dispersions
$\sigma$ assuming a power-law scaling with stellar mass,
\begin{equation}\label{eq:mason}
\log{\sigma} = \log{\sigma_0} + \beta_\sigma(\log{\msalp} - 11.5), 
\end{equation}
with $\log{\sigma_0}=2.48$ and $\beta_\sigma=0.20$ and a scatter of
$0.05$ in $\log{\sigma}$. These values are such that the evolved
  population of mock galaxies reproduce the $\msalp-\sigma$ relation
  measured at $z\sim0$ by \citet[][see
    \Sref{ssec:drymerger}]{Aug++10}.

\subsection{The effective IMF}
\label{ssect:imfform}

The effectve IMF of massive galaxies has been shown to correlate
with stellar velocity dispersion \citep[e.g.][]{Tre++10, CvD12,
  Cap++12, LaB++13, Spi++14, Pos++15}, stellar mass \citep{Aug++10b, Son++15},
 and stellar-mass density
\citep{Spi++15}.  In light of these observations, we adopt a relation
of the following form for assigning the IMF to our model galaxies at
$z=2$:
\begin{equation}\label{eq:imfform}
\log{\aimf} = a_*(\log{\msalp} - \mu_*) + a_\sigma(\log{\sigma} - \mu_\sigma) + b,
\end{equation}
where $\mu_*$ and $\mu_\sigma$ are the average logarithms of stellar mass and velocity dispersion of the sample.
We then choose two different sets of values for the parameters $a_*$, $a_\sigma$ and $b$. In the first prescription, labeled ``$M_*$ model'', we set $a_*=0.26$, $a_\sigma=0$ and $b=0.21$.
For the second prescription, labeled ``$\sigma$ model'', we set $a_*=0$, $a_\sigma=1.20$ and $b=0.03$.
In other terms, the first model is a power-law relation between IMF and stellar mass, with no residual dependence on $\sigma$, while vice-versa the IMF of the second model is set uniquely by the velocity dispersion.
In each model, the values of the parameters have been tuned to approximately reproduce the low-z IMF measurements from S15. Details on these measurements are given in \Sref{sect:obs}.

These prescriptions are purely empirically motivated. While there have been attempts at predicting the stellar IMF from first principles in terms of the global properties of a galaxy \citep[e.g.][]{Kru11,Hop12}, implementing such models would require us to make additional assumptions on parameters of the star forming gas such as the pressure and the turbulence Mach number.
Given the current uncertainty on the true mechanism determining the IMF, the benefits of employing such theoretically motivated recipes are modest and we therefore limit our model to simpler empirical recipes.

\subsection{The dry merger evolution}
\label{ssec:drymerger}

Our central galaxies are evolved to $z\sim0$ using the dry merger evolution model developed by \citet{Nip++12}, with a few modifications.
The method can be summarized as follows.
For each central galaxy we compute the evolution in its halo mass  using the following expression from \citet{FMB10}, which is derived from the Millennium I and II simulations:
\begin{equation}
\frac{\mathrm{d}\ln{\mhalo}}{\mathrm{d}z} = - \frac{\dot{M}_0}{10^{12}M_\odot H_0}\frac{1 + fz}{1 + z}\left(\frac{\mhalo}{10^{12}M_\odot}\right)^{g-1},
\end{equation}
with $\dot{M}_0 = 46.1 M_\odot \rm{\, yr}^{-1}$, $f=1.11$ and $g=1.1$.
In this step we are assuming that the growth history of each halo in our sample is the same as the average growth history for halos of the same mass.
We separate the growth in halo mass into smooth accretion and growth due to mergers. We estimate the latter with the following expression from \citet{FMB10},
\begin{equation}\label{eq:nmerg}
\frac{\mathrm{d}^2N_{\mathrm{merg}}}{\mathrm{d}z \mathrm{d}\xi}(z, \xi, \mhalo) = A\left(\frac{\mhalo}{10^{12}M_\odot}\right)^\alpha \xi ^\beta \exp{\left[\left(\frac{\xi}{\tilde{\xi}}\right)^\gamma\right]}(1 + z)^{\eta'},
\end{equation}
which describes the number of mergers between halos of mass ratio
$\xi$ per unit redshift interval.  Following \citet{FMB10}, we assume
$A=0.0104$, $\tilde{\xi}=9.72\times10^{-3}$, $\alpha=0.133$,
$\beta=-1.995$, $\gamma=0.263$ and $\eta'=0.0993$.  At each timestep
$dz$, the change in halo mass due to mergers is then the integral
over mergers of different halo mass ratios $\xi$,
\begin{equation}
\left[\frac{\mathrm{d}\mhalo}{\mathrm{d}z}\right]_{\mathrm{merg}} = -\int_{\ximin}^1 \mhalo\xi \frac{\mathrm{d}^2N_{\mathrm{merg}}}{\mathrm{d}z \mathrm{d}\xi} d\xi.
\end{equation}
We are assuming that only mergers with mass ratio larger than a
minimum value $\ximin$ contribute. This minimum value is set by the
merging time-scale: mergers with very small values of $\xi$ have
merging time-scales longer than the age of the Universe and therefore
cannot contribute to the accreted mass. We fixed $\ximin=0.03$
following \citet{Nip++12}. We tested for the impact of this choice on
our study by varying $\ximin$ between $0.01$ and $0.1$, finding
minimal differences in all the results.

We then calculate the growth in Salpeter stellar mass assuming that
each accreted halo is associated with a Salpeter stellar mass
  given by the (redshift dependent) SHMR of \citet{Lea++12}, and that this stellar
mass is also accreted by the central galaxy:
\begin{equation}
\frac{\mathrm{d}\msalp}{\mathrm{d}z} = -\int_{\ximin}^1 \mathcal{R}_{*,h}(\xi \mhalo, z)\xi\mhalo\frac{\mathrm{d}^2N_{\mathrm{merg}}}{\mathrm{d}z \mathrm{d}\xi} d\xi,
\end{equation}
where
$\mathcal{R}_{*,h}$ is the SHMR. With this equation we are assuming
that each merger event brings in exactly the average amount of stellar
mass for its halo mass: in other words we are neglecting scatter
  in the SHMR of the accreted satellites.

\begin{figure}
 \includegraphics[width=\columnwidth]{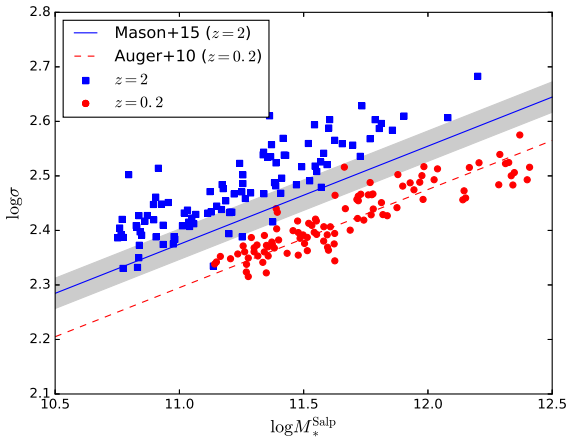}
 \caption{ Velocity dispersion as a function of stellar mass of
     the ``$M_*$ model'' mock sample (the ``$\sigma$ sample'' behaves
     similarly), at $z=0.2$ ({\em red circles}) and $z=2$ ({\em blue
     squares}).  {\em Dashed line:} best-fit $\msalp-\sigma$ relation
   measured by \citet{Aug++10} on a sample of strong lens early-type
   galaxies.  {\em Blue line:} $\msalp-\sigma$ relation
   at $z=2$ inferred by \citet{Mas++15}. 
   The $68\%$
   confidence region is marked by the shaded region.}
 \label{fig:vdisp}
\end{figure}

Each merger, in addition to increasing the mass of the galaxy,
contributes to modify its velocity dispersion. Using the virial
theorem under the assumption of parabolic orbits, \citet{NJO09} showed
that the ratio between the {\em virial} velocity dispersion after and
before a merger event of mass ratio $\xi$ is given by
\begin{equation}
\frac{\sigma_{v,f}^2}{\sigma_{v,i}^2} = \frac{1 + \xi\epsilon}{1 + \xi},
\end{equation}
where $\epsilon = \sigma_{v,a}^2/\sigma_{v,i}^2$ is the ratio between
the velocity dispersion of the accreted object and that of the central
galaxy, squared.  When $\epsilon<1$ (as expected when $\xi<1$),
  $\sigma_{v,f}<\sigma_{v,i}$, so the velocity dispersion is expected
  to decrease as a consequence of a gas-poor, parabolic minor
  merger. We make the assumption that, for both the central and the
  satellite galaxies, the stellar velocity dispersion is proportional
  to their virial velocity dispersion $\sigma \propto \sigma_v$, so we
  can write the change in stellar central velocity dispersion due to
  mergers as
\begin{equation}
\frac{\mathrm{d}^2\sigma^2}{\mathrm{d}z\mathrm{d}\xi} = \sigma^2\left[1-\frac{1 + \xi\epsilon}{1 + \xi}\right]\frac{\mathrm{d}^2N_{\mathrm{merg}}}{\mathrm{d}z \mathrm{d}\xi}.
\end{equation}
We
describe the velocity dispersion of the satellites with a relation
between stellar mass and $\sigma$ of the form given by
\Eref{eq:mason}, with no scatter. The values of the coefficients
$\sigma_0$ and $\beta_\sigma$ are determined by fitting
\Eref{eq:mason} to the $\msalp-\sigma$ relation of the central
galaxies, and are updated at each timestep as the centrals evolve.  In
\Fref{fig:vdisp} we show the distribution in velocity dispersion of
the mock galaxies as a function of stellar mass, at $z=0.2$ and
$z=2$. We compare these values with the low-$z$ observations from
\citet{Aug++10} and the constraints at $z=2$ inferred by
\citet{Mas++15}. Our model tends to overpredict the velocity
  dispersions for a given stellar mass with respect to the values
  measured by \citet{Mas++15}. The effects of this possible
discrepancy on the results of this work are discussed in the next 
sections.

Finally, we need to describe the growth in true stellar mass for our centrals, which in turn requires to define the IMF of the accreted satellites.
Similarly to the velocity dispersion case, we assign the IMF of satellites using \Eref{eq:imfform}, where the coefficients are determined at each timestep by fitting the distribution of IMF as a function of stellar mass and velocity dispersion of the centrals.

We do not keep track of the spatial distribution of stars within our galaxies. Although observations could in principle detect spatial variations in IMF within a galaxy \citep{Mar++15}, provided that metallicity gradients are accurately accounted for \citep{MLM16}, and put interesting constraints on IMF differences between the in-situ population and that of the accreted stars, that is left for future work.

\section{Reference IMF measurements}\label{sect:obs}

The main goal of this work is to make predictions on the distribution
of the IMF in massive galaxies as a function of redshift, using
low-redshift observations as a reference point. 
To ensure self-consistency when comparing with wesults at high redshift, we choose the IMF measurements from \citet[][hereafter S15]{Son++15} to
be such reference point.
In this Section we summarize the main
results of that work and carry out some additional analysis needed in
order to calibrate our models to the observations.

S15 measured the effective stellar IMF for a sample of 80
early-type galaxies using strong lensing in combination with velocity
dispersion data to constrain the stellar mass $\mtrue$ of each object.
Measurements of $\aimf$ for individual objects are very noisy. To cope
with the low signal-to-noise of individual measurements S15 combined
them with a hierarchical Bayesian inference approach, in which they
modeled the distribution of IMF normalization of the population of
massive galaxies as a Gaussian and allowed the mean of such Gaussian
to be a function of redshift, stellar mass and projected stellar mass
density.  The analysis yielded the detection of a positive trend
between the effective IMF and stellar mass and no detection of a
dependence on redshift or stellar mass density.

The S15 analysis did not explore dependences between velocity dispersion and IMF, which we consider in our model.
Moreover, the relations between stellar mass and IMF explored by S15 were based on $\mtrue$, instead of $\msalp$.
For this reason we re-analyse here the S15 sample and fit for a dependence of the IMF on $\msalp$ and $\sigma$, as well as redshift. We then describe the distribution in effective IMF of the S15 sample of galaxies as a Gaussian with mean
\begin{equation}\label{eq:sl2sfit}
a_z(z - 0.3) + a_*(\log{\msalp} - 11.5) + a_\sigma(\log{\sigma} - 2.4) + \log{\alpha_{\mathrm{IMF},0}}
\end{equation}
and dispersion $s$. 
The pivot points $z=0.3$, $\log{\msalp}=11.5$ and $\log{\sigma}=2.4$ are very close to the median values of the corresponding quantities in the S15 sample. This allows us to minimize the covariance in the inference of the paramteres $a_z$, $a_*$ and $a_\sigma$.
We fit for $a_z$,
$a_*$, $a_\sigma$, $\alpha_{\mathrm{IMF},0}$ and $s$ using a similar
hierarchical Bayesian inference method as the one used by S15.
Details of the fitting procedure are given in the Appendix.  The main
assumptions of this model are: a fixed Navarro Frenk and White
\citep[][NFW]{NFW97} shape for the dark-matter distribution, a de Vaucouleurs \citep{deV48} profile with spatially constant
mass to-light-ratio and isotropic velocity distribution
for the stars.  

\begin{figure*}
 \includegraphics[width=\textwidth]{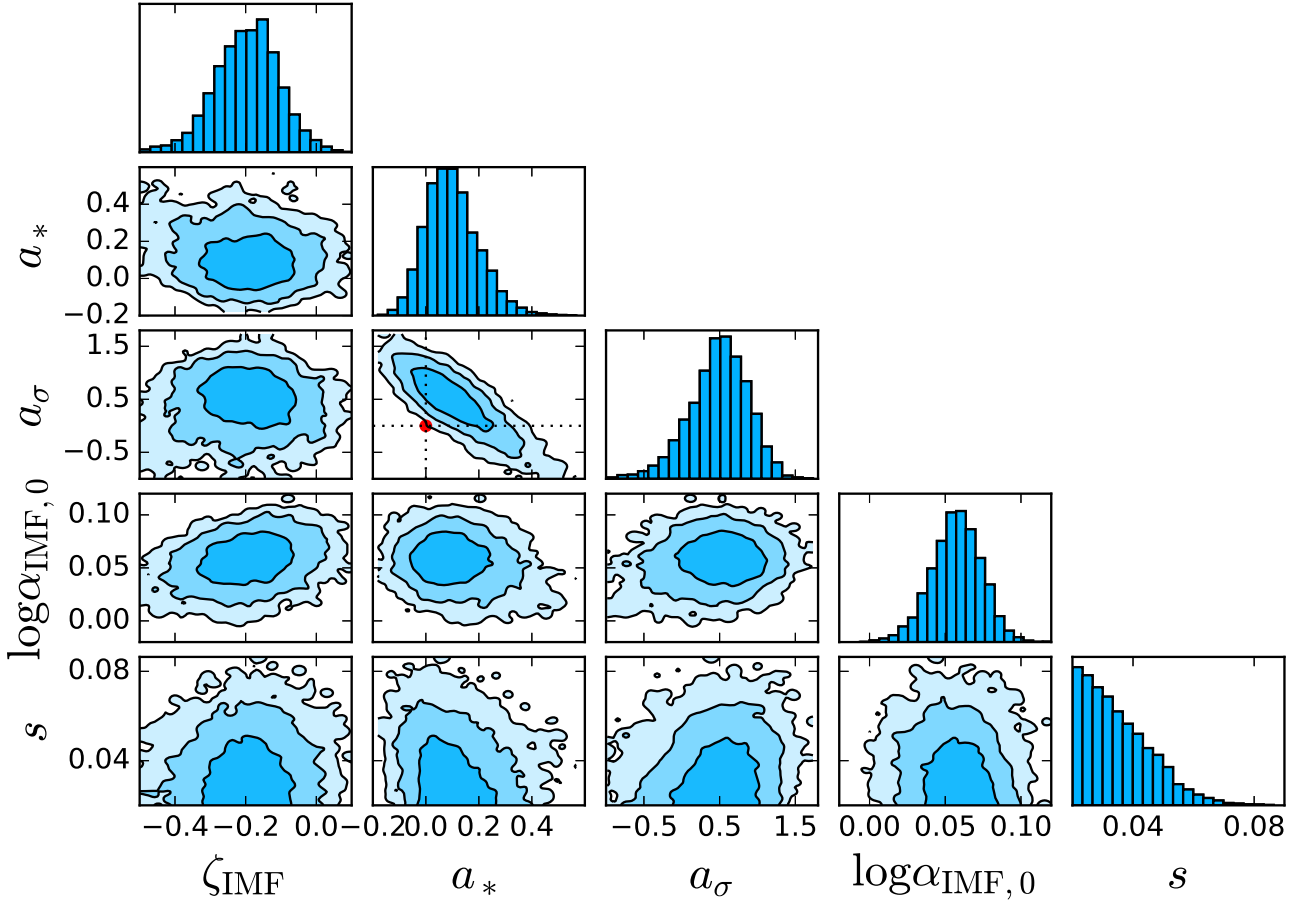}
 \caption{Posterior probability distribution of parameters describing
   the effective IMF of early-type galaxies, modeled as a Gaussian with mean given by \Eref{eq:sl2sfit} and dispersion $s$ and fitted to 
   lensing and stellar kinematics observations of the S15 sample
   . Different levels represent the 68\%, 95\% and
   99.7\% enclosed probability regions.
The red dot marks the values $a_*=0$ and $a_\sigma=0$, corresponding to a Universal IMF.
}
 \label{fig:sl2scorner}
\end{figure*}

In \Fref{fig:sl2scorner} we plot the posterior probability distribution for the parameters describing the IMF, marginalized over the other model parameters (which include parameters describing the dark-matter distribution).
Consistently with the S15 analysis, we find an average IMF
normalization slightly heavier than a Salpeter IMF
($\log{\alpha_{\mathrm{IMF},0}} = 0.06\pm0.02$). There is a strong
degeneracy between the parameters $a_*$ and $a_\sigma$, meaning that
the data can be described either with a dependence of the IMF on
stellar mass or on velocity dispersion. This degeneracy is a
consequence of the tight correlation between stellar mass and velocity
dispersion in early-type galaxies.  Nevertheless the data, under the
assumptions specified above and discussed in
Appendix~\ref{sect:appendix}, exclude a Universal IMF at the
$3-\sigma$ level and show a trend of increasing IMF normalization with
increasing mass and/or velocity dispersion.

\begin{figure*}
 \includegraphics[width=\textwidth]{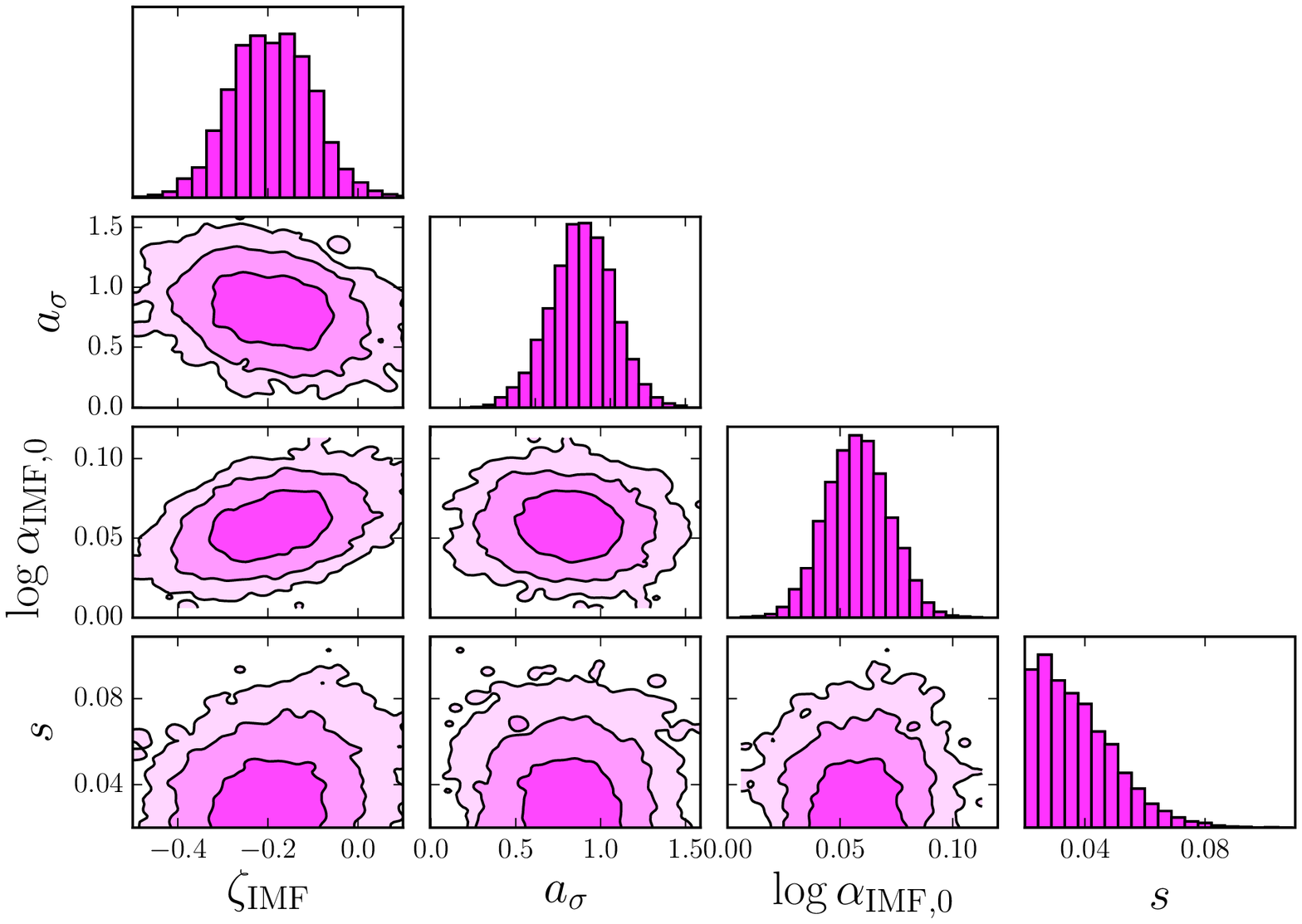}
 \caption{
Posterior probability distribution of parameters describing
   the effective IMF of early-type galaxies, modeled as a Gaussian with mean given by \Eref{eq:sigmaonlyfit} and dispersion $s$ and fitted to 
   lensing and stellar kinematics observations of the S15 sample
   . Different levels represent the 68\%, 95\% and
   99.7\% enclosed probability regions.
}
 \label{fig:sigmaonlycorner}
\end{figure*}

Given that 1) our inference on the parameter $a_*$ is consistent with zero, 2) a description of the galaxy population in which the IMF depends simultaneously on velocity dispersion and stellar mass is not of trivial interpretation and 3) a large number of works in the literature focus on trends of the IMF with velocity dispersion alone, we repeat the analysis by fixing $a_*=0$. 
In this simpler description of the data, the IMF mismatch parameter is then distributed as a Gaussian with mean given by
\begin{equation}\label{eq:sigmaonlyfit}
a_z(z - 0.3) + a_\sigma(\log{\sigma} - 2.4) + \log{\alpha_{\mathrm{IMF},0}}
\end{equation}
and dispersion $s$.
The posterior probability distribution of the parameters of this model is plotted in \Fref{fig:sigmaonlycorner}. 
%


\section{Results}\label{sect:results}

\subsection{Evolution of the $\aimf$-$\msalp$ and $\aimf$-$\sigma$ relations}

\begin{figure*}
 \includegraphics[width=\textwidth]{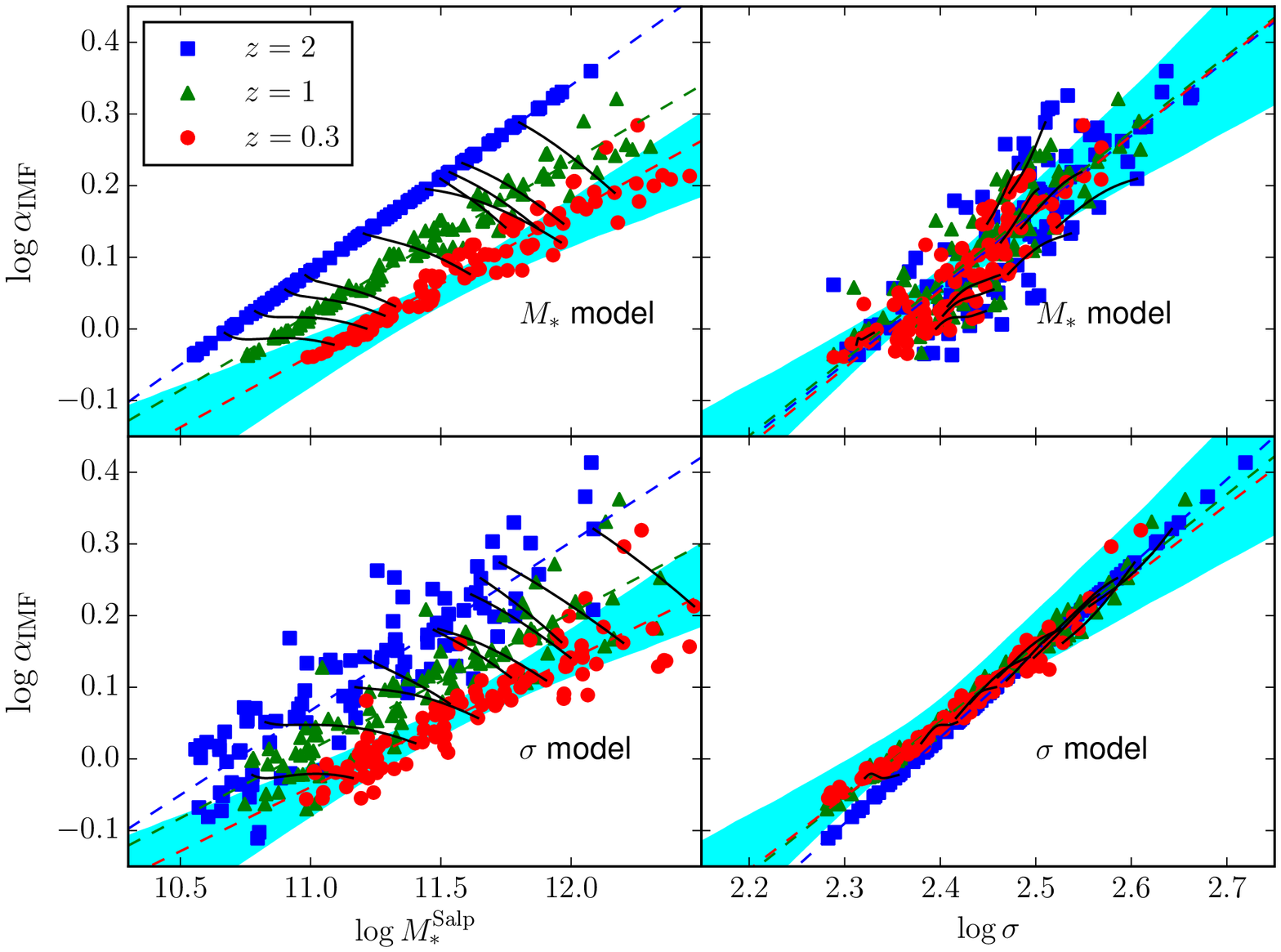}
 \caption{ IMF mismatch parameter $\aimf$ as a function of stellar
   mass ({\em left panels}) and central velocity dispersion ({\em
     right panels}) for the ``$M_*$ model'' ({\em top panels}) and
   ``$\sigma$ model'' ({\em bottom panels}) IMF recipes, at $z=2$,
   $z=1$ and $z=0.3$.  {\em Solid lines:} Evolutionary tracks between
   $z=2$ and $z=0.3$ of a few representative galaxies.  {\em Dashed
     lines:} fits of \Eref{eq:mstarmodel} and \Eref{eq:sigmamodel} to the data in each redshift snapshot. {\em
     Shaded regions:} Observational constraints from \Sref{sect:obs},
   obtained using data from S15 at $z=0.3$.   }
 \label{fig:snap}
\end{figure*}

We evolved our mock population of galaxies from $z=2$ to $z=0.3$, the
median redshift of the S15 sample and the redshift at which lensing constraints on the IMF are most robust. In \Fref{fig:snap} we
plot the IMF mismatch parameter for our mock sample of galaxies at
three different redshifts ($z=2$, $z=1$ and $z=0.3$), as a function of
stellar mass and velocity dispersion, for each model.
At each redshift snapshot and for each model we fit for a power-law dependence of the IMF on stellar mass 
\begin{equation}\label{eq:mstarmodel}
\log{\aimf} = c_*(\log{\msalp} - 11.5) + d_*
\end{equation}
and velocity dispersion
\begin{equation}\label{eq:sigmamodel}
\log{\aimf} = c_\sigma(\log{\sigma} - 2.4) + d_\sigma.
\end{equation}
Best-fit curves are overplotted in \Fref{fig:snap} and values of the parameters are reported in \Tref{tab:oneparfit}.
\begin{table}
 \caption{Best-fit parameters of \Eref{eq:mstarmodel} and \Eref{eq:sigmamodel} to the mock data shown in \Fref{fig:snap}. The third number in each set represents the residual scatter around the best fit relation.}
 \label{tab:oneparfit}
 \begin{tabular}{lcc}
 \hline
 Data set & Eq. \ref{eq:mstarmodel} fit & Eq. \ref{eq:sigmamodel} fit\\
 & $(c_*, d_*, s_*)$ & $(c_\sigma, d_\sigma, s_\sigma)$ \\
 \hline
 ''$M_*$ model'' & (0.26, 0.21, 0.00) & (1.06, 0.06, 0.06)\\
''$M_*$ model'' & (0.21, 0.13, 0.02) & (1.06, 0.06, 0.04)\\
''$M_*$ model'' & (0.20, 0.06, 0.02) & (1.08, 0.05, 0.04)\\
''$\sigma$ model'' & (0.24, 0.19, 0.05) & (1.20, 0.03, 0.00)\\
''$\sigma$ model'' & (0.19, 0.11, 0.04) & (1.06, 0.05, 0.01)\\
''$\sigma$ model'' & (0.18, 0.05, 0.03) & (1.02, 0.05, 0.01)\\

 \hline
 \end{tabular}
\end{table}

As can be seen by comparing the left hand panels in \Fref{fig:snap}
with the right hand panels, correlations of the effective IMF
  with stellar mass correspond to similar correlations with velocity
  dispersion and vice-versa. This is again a consequence of the
  relatively tight correlation between galaxy stellar mass and
velocity dispersion in the mock sample (and in observations;
\Fref{fig:vdisp}).  By looking at the redshift evolution of our mock
samples we can see that the IMF normalization of individual objects
decreases with time. This can be easily understood: central galaxies
grow by merging with smaller objects which, by construction, have a
lighter IMF and therefore bring the effective IMF of the post-merger
galaxy towards smaller values.  Focusing on the left-hand side of
\Fref{fig:snap} one can also notice how the most massive galaxies
experience the largest decrease in IMF normalization. As a result, the
correlation between stellar mass and effective IMF becomes shallower
with time (the coefficient $c_*$ in \Eref{eq:mstarmodel} decreases with time). The reason for this tilt will be discussed in
\Sref{sect:discuss}.  
A similar trend is seen for the correlation between $\sigma$ and $\aimf$ in the ``$\sigma$ model'' (bottom right panel of \Fref{fig:snap}), becoming shallower with time. In contrast, for the ``$M_*$ model'' the original correlation between velocity dispersion and IMF is roughly preserved by dry mergers.


While the mean trends between stellar mass or velocity dispersion and
stellar IMF evolve very similarly for both models examined here, the
two models differ in terms of scatter around these correlations.  The
``$M_*$ model'' is initialized with zero scatter around the
$\msalp-\aimf$ relation, but by $z=1$ a substantial scatter is
introduced.  For the ``$\sigma$ model'' however the relation between
$\aimf$ and $\sigma$ in place at $z=2$ remains remarkably tight down
to $z=0.3$.  This is a consequence of the different direction of the
evolutionary tracks in $\msalp-\aimf$ and $\sigma-\aimf$ space. Objects
evolve roughly perpendicular to the $\msalp-\aimf$ relation, quickly
introducing scatter, while as discussed above the $\sigma-\aimf$
relation is mostly preserved due to the coincidence between direction
of evolution and correlation in $\sigma-\aimf$ space.  This prediction
depends critically on our model for the evolution of the velocity
dispersion with dry mergers, based on the assumption of parabolic
orbits for the accreted satellites, which predicts steadily
  declining values of $\sigma$ with time for individual galaxies.
Dark matter-only cosmological simulations show that the parabolic
orbit approximation tends on average to underestimate the post-merger
velocity dispersion \citep{Pos++14}.  For instance, in the simple
  case in which $\sigma$ remains constant after each merger, objects
  would move vertically in the $\sigma-\aimf$ plane, qualitatively
  changing the right hand panels of \Fref{fig:snap}.


\subsection{Evolution of the mock galaxies in the $\aimf$-$\msalp$-$\sigma$ space}

\begin{figure}
 \includegraphics[width=\columnwidth]{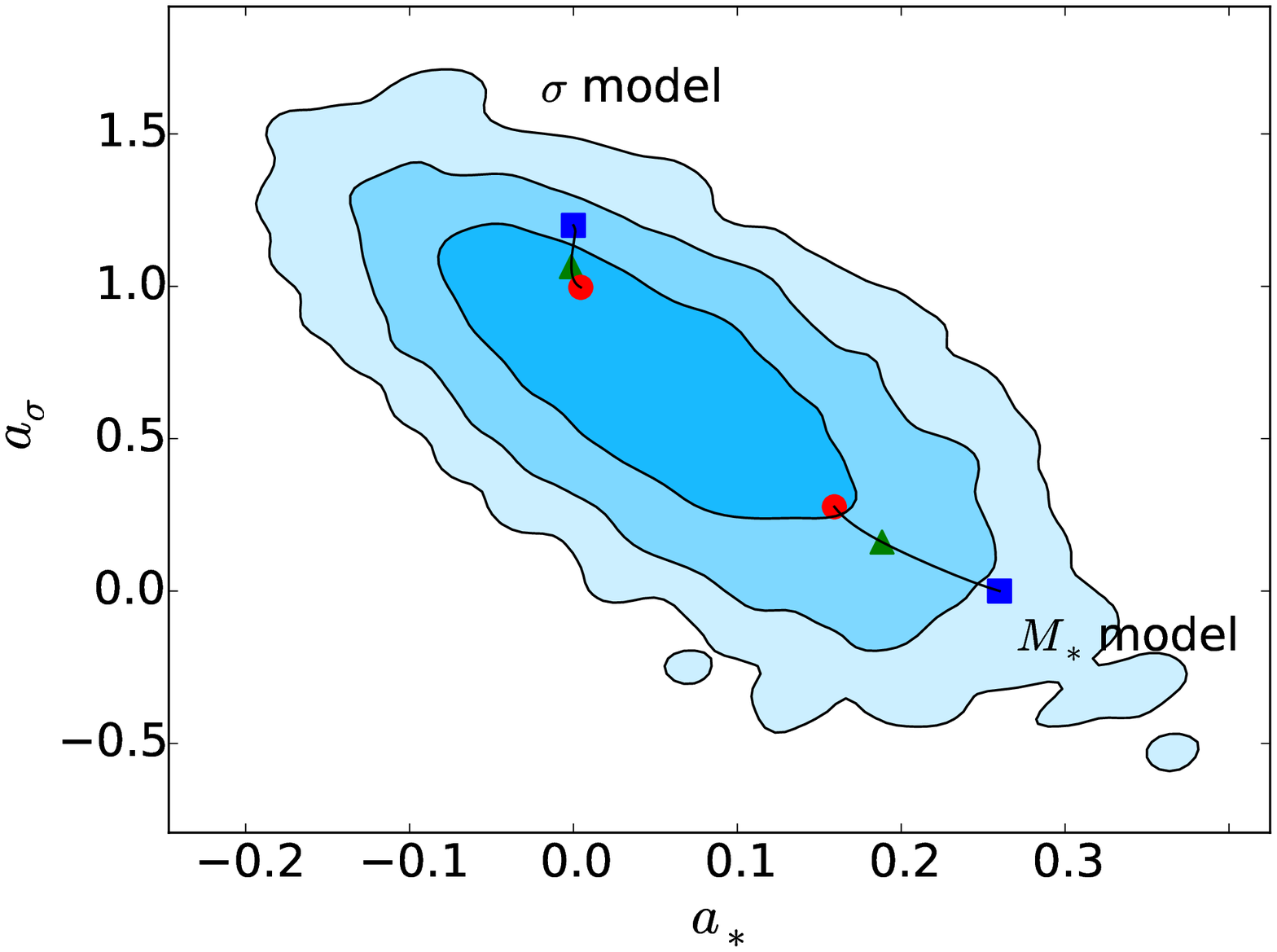}
 \caption{Coefficients $a_*$ and $a_\sigma$, describing the dependence
   of the effective IMF on stellar mass and velocity dispersion,
   obtained by fitting \Eref{eq:imfform} to mock data generated with
   the ``$M_*$ model'' and ``$\sigma$ model'' prescriptions at each
   timestep.  The shaded region marks the observational constraint at
   $z=0.3$ obtained in \Sref{sect:obs} from S15 data. 
   Different contours mark regions of 68\%, 95\% and 99.7\% enclosed probability.
}
 \label{fig:tracks}
\end{figure}

\Fref{fig:snap} only shows how the parameters $\msalp$ and $\sigma$
correlate {\em individually} with the effective IMF.  We now examine
simultaneous dependences of the IMF on stellar mass and velocity
dispersion, that is how $\aimf$ scales with $\msalp$ at fixed $\sigma$
and viceversa.  For each mock sample at each timestep we determine the
average of the logarithm of the stellar mass and velocity dispersion,
$\mu_*(z)$ and $\mu_\sigma(z)$, then fit \Eref{eq:imfform} to the IMF
distribution and plot the measured coefficients $a_*$ and $a_\sigma$
in \Fref{fig:tracks}.

%

The starting ($z=2$) point in the $a_*-a_\sigma$ space is by
construction very different for the two mock sets.  For the ``$M_*$
model'' the initial dependence on stellar mass is in part converted
into a dependence on velocity dispersion as a result of dry mergers.
By contrast, the coefficients describing the $\sigma$ model at $z=0.3$
are very close to the $z=2$ values. The initial dependence of the IMF
on velocity dispersion becomes slightly weaker while no significant
dependence on stellar mass (at fixed velocity dispersion) is
introduced. This is a result of dry mergers preserving the tight
correlation between $\sigma$ and $\aimf$, apparent in the bottom right
panel of \Fref{fig:snap}.  As discussed above, this prediction relies
in turn on the assumed evolution of the velocity dispersion. If we
allow for a milder decline in velocity dispersion with dry mergers we
observe a more significant mixing between stellar mass and velocity
dispersion dependece of the IMF, the red dot corresponding to the
``$\sigma$ model'' in \Fref{fig:tracks} approaching the corresponding
dot of the ``$M_*$ model''.

\subsection{Average IMF normalization as a function of redshift at fixed velocity dispersion}

Finally we examine how the IMF normalization of the mock sample at fixed galaxy properties evolves in time and compare it with the S15 measurements.
In \Sref{sect:obs} we presented two different descriptions of the IMF distribution: one in which $\aimf$ scales with redshift, stellar mass and velocity dispersion (equation \ref{eq:sl2sfit}) and a simpler one in which the dependence on stellar mass is ignored (equation \ref{eq:sigmaonlyfit}).
Since, as shown in \Fref{fig:tracks}, the S15 data is unable to put interesting constraints on the separate dependence of the IMF on $\msalp$ and $\sigma$, we adopt the latter description from now on and consider for simplicity only dependences of $\aimf$ on velocity dispersion and redshift.

\begin{figure*}
 \includegraphics[width=\textwidth]{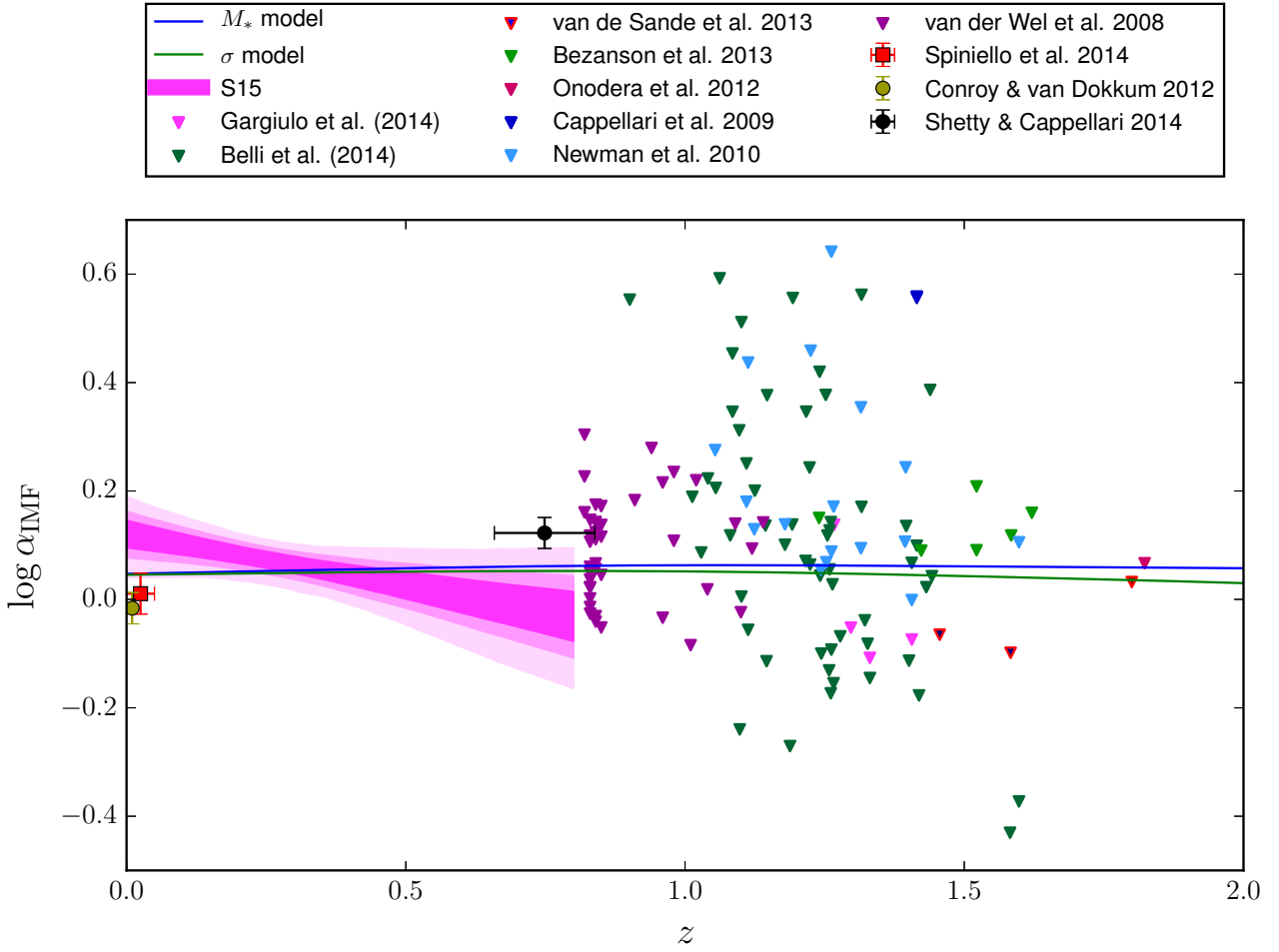}
 \caption{ {\em Solid lines:} Average effective IMF
   normalization for galaxies at 
   $\log{\sigma}=2.4$ of the mock samples, parameter $d_\sigma$ in 
   \Eref{eq:sigmamodel}, as a function of redshift.  {\em Shaded
     regions:} Observational constraints from the analysis of S15 data
   carried out in \Sref{sect:obs}, obtained by evaluating
   \Eref{eq:sigmaonlyfit} at $\log{\sigma}=2.4$. Different levels mark the 68\%,
   95\% and 99.7\% enclosed probability regions.  {\em Points with
     error bars:} Average IMF normalization at 
   $\log{\sigma}=2.4$ inferred by \citet{CvD12}, \citet{Spi++14} and
   \citet{S+C14}. Horizontal error bars indicate the range in redshift
   of the galaxies within the samples.  {\em Triangles:}
   upper limit estimates of $\aimf$ from isotropic mass-follows-light spherical
   Jeans dynamical models applied to velocity dispersion measurements
   from the literature, corrected to the pivot point assuming the value
   of $a_\sigma$ measured in the S15 analysis. The triangle 
   marks the $1-\sigma$ (84 percentile) limit.
   }
 \label{fig:bevol}
\end{figure*}

In \Fref{fig:bevol} we plot the mean IMF normalization for galaxies at $\log{\sigma}=2.4$ of the mock populations together with the value measured in the analysis of the S15 data.
For the model curves, this is given by parameter $d_\sigma$ of \Eref{eq:sigmamodel}, fitted for at each timestep. For the S15 measurements, the observational band is obtained by evaluating \Eref{eq:sigmaonlyfit} at $\log{\sigma}=2.4$ at each redshift between $z=0$ and $z=0.8$ (there are no objects at $z>0.8$ in the S15 sample).  
We see a $2-\sigma$ discrepancy between both models and the
S15 constraints. 

We then compare our models with several sets of measurements from the
literature.  
\citet{CvD12} constrained the IMF of a sample of nearby massive
ellipticals by fitting stellar population models to spectral indices
sensitive to the abundance of low-mass stars.  They found a trend
between IMF slope, and normalization, and stellar velocity
dispersion. We took their measurements of the IMF normalization,
converted them to our definition of $\aimf$, then fitted
for a power-law dependence between $\aimf$ and $\sigma$ and plotted
the value inferred at $\log{\sigma}=2.4$ in \Fref{fig:bevol}.  Using a
similar technique, \citet{Spi++14} constrained the IMF of low redshift
early-type galaxies by fitting stellar population models to a large
sample of Sloan Digital Sky Survey spectra.  They measured a positive
trend between IMF normalization and galaxy velocity dispersion. In
\Fref{fig:bevol} we plot the value of $\aimf$ inferred by
\citet{Spi++14} for galaxies at $\log{\sigma} = 2.4$.

These are all low-redshift measurements, plotted with the purpose of
estimating the systematic uncertainty in the determination of $\aimf$
with different techniques.  At higher redshift, \citet{S+C14}
constrained the IMF of 68 galaxies at $z\sim0.75$ from dynamical
modeling of single aperture stellar kinematics data. 
Their analysis assumes that the total
density profile follows the light distribution, neglecting the
contribution of dark matter.

Although \citet{S+C14} report values of the IMF normalization and
velocity dispersion for each object, we cannot use this information
directly to fit for a trend between $\aimf$ and $\sigma$ because in
dynamical analysis, unlike in the spectral fitting technique, the
measurement uncertainty on the velocity dispersion is strongly
correlated with that on the IMF.  An accurate fit of
$\aimf$ versus $\sigma$ would then require a full knowledge of the
measurement uncertainty in the $\sigma-\aimf$ plane, which, unlike for the S15 sample, is not
available.  We then fix the slope of the $\sigma-\aimf$ relation to
the value of $a_\sigma$ measured in the analysis of the S15 data and fit for the
normalization of the IMF at $\log{\sigma}=2.4$ in the \citet{S+C14}
sample.  The resulting value of $\aimf$ in slight tension with both the ``$M_*$ model'' and the ``$\sigma$ model'', though this
tension is likely to disappear once the contribution from dark matter
is taken into account.

At redshift higher than $z\sim0.8$ there are no robust measurements of
the IMF normalization.  However we can put upper limits on the stellar
mass, and the IMF, of galaxies with measured velocity dispersion using
dynamical arguments.  We can write the virial theorem for a stationary
stellar system,
\begin{equation}\label{eq:virial}
K_V\sigma_{e2}^2 = G\frac{M}{R_e},
\end{equation}
where $M$ is the total (stellar plus dark-matter) mass, $R_e$ the
projected half-light radius, $\sigma_{e2}$ the luminosity-weighted
line of sight velocity dispersion within a circular aperture of radius
$R_e/2$, and $K_V$ is a coefficient that takes into account the
density profile of mass and light, geometrical effects and the
distribution of stellar orbits.  In the limit in which all the
  mass is in stars (no dark matter) $M=\mtrue$, so the virial
  coefficient $K_V$ can be used to put an upper limit on the stellar
  mass.

We then collect measurements of $\sigma_{e2}$ in the redshift range
$0.8 < z < 2$ from the literature. Following \citet{Mas++15} we
consider measurements from \citet{vdW++08}, \citet{Cap++09},
\citet{New++10}, \citet{Ono++12}, \citet{Bez++13}, \citet{vdS++13}, as
well as measurements from \citet{Gar++15}.  We assume isotropic
  stellar velocity distribution, consistently with the analysis of the
  S15 data, and use the spherical Jeans equation to calculate $K_V$
  for a mass-follows-light S\'{e}rsic \citep{Ser68} profile with structural parameters
  given by the observed values. 
The adopted values of $K_V$ range from $3.19$ (corresponding to a S\'{e}rsic index $n=8.0$) to $7.18$ (corresponding to $n=1.1$).
We
solve for $M=\mtrue$ in \Eref{eq:virial} and divide the stellar mass
by the value obtained from stellar population synthesis assuming a
Salpeter IMF to obtain $\aimf$.  
Finally, we correct these
measurements to the pivot point 
$\log{\sigma}=2.4$ using the values of $a_\sigma$ measured
in the analysis of S15 data.  $1-\sigma$ upper limits ($84\%$ enclosed
probability) on $\log{\aimf}$ for individual galaxies are plotted in
\Fref{fig:bevol}.  39\% (34\%) of the objects have $1-\sigma$ upper
limits below the ``$M_*$ model'' (``$\sigma$ model''), indicating
tension between both models and observations.  Part of this tension
could simply be due to intrinsic scatter in the distribution of IMF in
the population of galaxies.  Although in \Sref{sect:obs} we
constrained the intrinsic scatter to be below $0.08$ in $\log{\aimf}$,
there are indications that the scatter in IMF might be larger 
\citep{S+L13}. 
Other possible causes for the discrepancy between models and observations are discussed in the next section.
\section{Discussion}\label{sect:discuss}

\subsection{Comparison of models with observations}

Our model predicts that the effective IMF $\aimf$ of massive
  early-type galaxies, at fixed velocity dispersion, should remain roughly constant with
  redshift. However, such a trend is not apparent
  from the few observational constraints on the IMF available at
  $z>0.3$: in fact some data, including the S15 measurements used to calibrate our model at $z=0.3$, suggest that $\aimf$ might be lower at higher
  $z$ \citep[see also][]{Tor++14}. 
 A possible
source for this tension is the treatment of the dark-matter component
in the analysis of the S15 data, in which we fixed the shape of
  the dark-matter density distribution to an NFW profile (see the
  Appendix). The dark-matter slope is however highly degenerate with
the IMF \citep[see e.g.][]{Aug++10}. In particular, if the
  dark-matter profile is getting steeper with time the IMF
  normalization inferred at low redshift would decrease, bringing the
  model in better agreement with the measurements. However such a
  hypothesis is difficult to verify with the current data.  

  Although
  the discrepancy between the models and the S15 measurements is
  within $3-\sigma$, a similar tension is seen in the comparison with
  dynamical estimates on the IMF at $z>1$.  Dark matter cannot be a
  source of bias in this case because its contribution is ignored in
  our treatment of high redshift measurements.

Another possible source of error that could be affecting the
measurements differently at different redshifts is the presence of
spatial gradients in the IMF.  As shown by \citet{New++15}, an IMF
that becomes less heavy at increasing distance from the center of a
galaxy would bias lensing and dynamical studies that assume a
spatially constant IMF (such as the observations presented here)
towards a globally heavier IMF. A radial gradient in IMF could be
created by dry mergers if the newly accreted material has a more
extended distribution than the preexisting stellar
distribution. Indeed, minor dry mergers are predicted to build up an
extended envelope of stars \citep{NJO09,Hop++10} which, in the context
of our model, would have a lighter IMF compared to the central
parts. Such an extended envelope of stars with lighter IMF would grow
in time, producing a stronger bias towards a heavier inferred IMF in
low redshift galaxies. \citet{Mar++15} and \citet{LaB++16} claimed a
detection of a spatial variation of the IMF, heavier in the center, in
three massive early-type galaxies.  Testing whether such a scenario
could be the cause of the apparent evolution in IMF would require a
much more complex model than the one explored in this work, and we
therefore leave it to future study.

Assuming that the tension between model and data is real and not the result of systematic errors or selection effects,
 our simple scenario based on dry mergers must be modified.  In
a previous work we showed how a very similar dry merger evolution
model to the one presented here is unable to reproduce both the size
evolution and the evolution in the slope of the density profile of
massive galaxies at $z<1$ \citep{SNT14}.  The model predicted density
slopes that become shallower with time, at odds with lensing
constraints. As a solution we proposed a scenario in which a small
fraction (about 10\%) of the accreted baryonic mass is in the form of
gas that falls to the center of the galaxy causing the density profile
to steepen and the velocity dispersion to increase as a result of
adiabatic contraction.  

Such a scenario would modify the predicted IMF
evolution in two ways: with the addition of a population of stars
formed in-situ and by modifying the evolution of the velocity
dispersion.  Contributing with less than 10\% of the final mass, the
newly formed stars would not change significantly the effective IMF of
a galaxy unless their IMF normalization differs by a factor of a few
with respect to that of the accreted satellites, which seems
unplausible. However, a central core of stars with a different IMF
from the preexisting population would create a radial gradient in the
IMF which, as discussed above, could be a source of bias for dynamical
measurements.  The increase in velocity dispersion caused by the
adiabatic contraction from the infall of gas would improve the
agreement with the observed evolution of the $\msalp-\sigma$ relation.
We have shown in \Fref{fig:vdisp} how our model tends to overpredict
the average velocity dispersion at $z=2$ with respect to observations,
meaning that a slower decline in velocity dispersion is required to
precisely match both $z=0$ and $z=2$ velocity dispersion measurements.
However, in a scenario in which the velocity dispersion evolves more
slowly the $\sigma-\aimf$ correlation will not be preserved, as
predicted in our purely dry-merger model (see right-hand panel of
\Fref{fig:snap}), but will become shallower with time. This means that
at fixed $\sigma$ the IMF will be heavier at larger $z$, increasing
the tension between model and $z>1$ data.

Finally, a possible solution to the discrepancy would be to allow for completely different IMF distributions between centrals and satellites. Our model assumes that the scaling relation between the effective IMF of accreted galaxies and their galaxy properties is the same as that of centrals, extrapolated to low masses.
If we allow, for instance, the IMF normalization of satellites to be on average heavier than that of centrals, individual galaxies would evolve towards a heavier effective IMF with time, in the direction suggested by the data. There are currently no robust observations through which we can test this hypothesis.

\subsection{IMF evolution and average merger mass ratio}

\begin{figure*}
 \includegraphics[width=\textwidth]{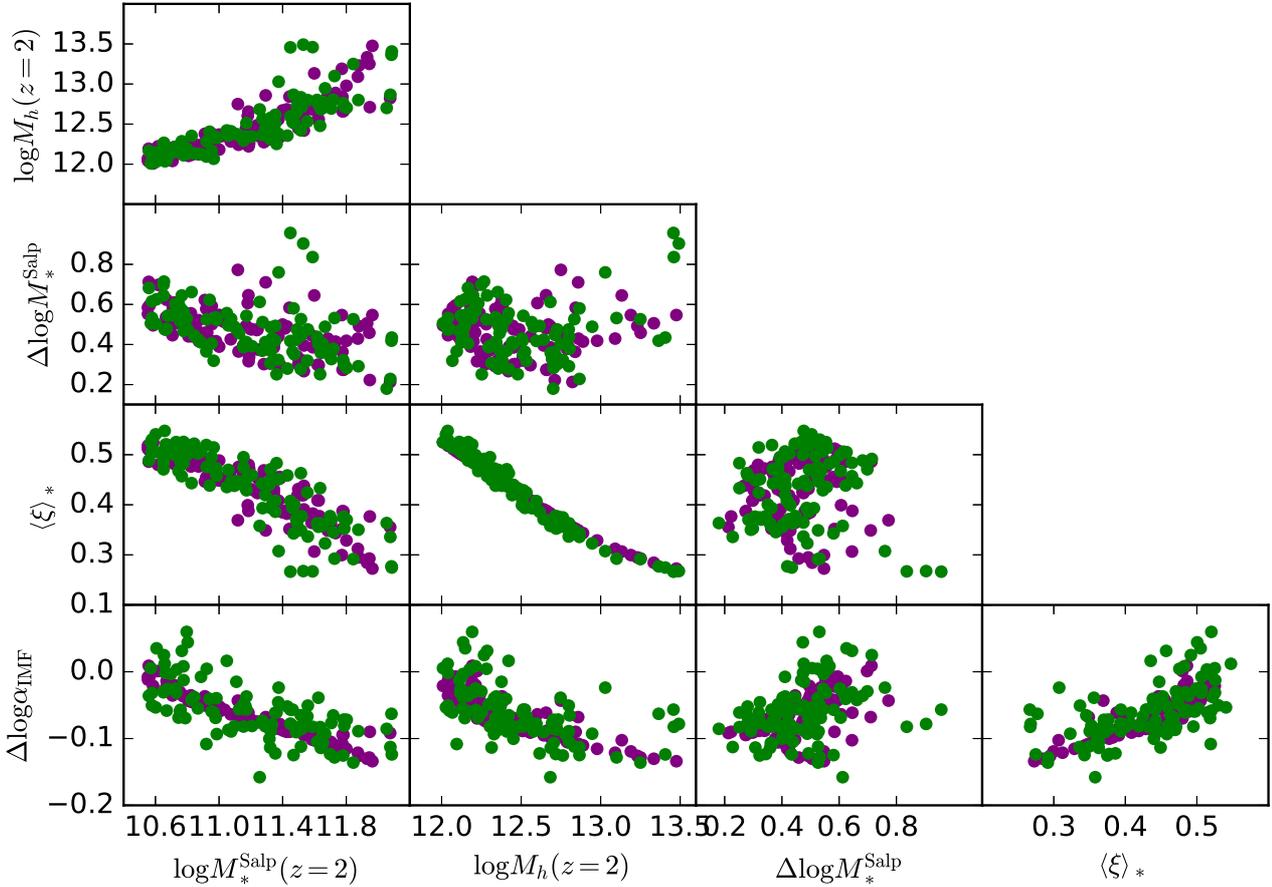}
 \caption{ 
   Distribution in stellar mass $\msalp$ and halo mass $\mhalo$ at $z=2$, fractional
   variation in stellar mass $\Delta\log{\msalp}$ and effective IMF $\Delta\log{\aimf}$ between $z=0$ and $z=2$, and stellar
   mass-weighted merger mass ratio $\left<\xi\right>_*$, as defined in \Eref{eq:xieff}.
   {\em Purple dots}: ``$M_*$ model''. {\em Green dots}: ``$\sigma$
   model''.  }
 \label{fig:cornerplot}
\end{figure*}

A second prediction of our model is that the deacrease of the IMF
mismatch parameter with time is stronger for the more massive
objects. This is shown by the evolution in the tilt of the
$\aimf-\msalp$ correlation with redshift seen in the left hand column
of \Fref{fig:snap}.  To understand the origin of this effect, in
\Fref{fig:cornerplot} we plot the fractional change in effective IMF
between $z=2$ and $z=0$,
$\Delta\log{\aimf}=\log{\aimf}(z=0)-\log{\aimf}(z=2)$ of our mock
galaxies together with key quantities such as the initial halo mass
and initial stellar mass and accreted stellar mass.
The effective IMF of individual galaxies in our sample generally
decreases with time ($\Delta\log{\aimf}<0$). This is easily
understood, as by definition central galaxies merge with less massive
systems that, in our model, have a lighter IMF.  In order to produce a
dependence of the IMF on mass that becomes shallower with time it is
necessary for the more massive systems to have a stronger decrease in
their IMF with time (more negative values of $\Delta\log{\aimf}$).  As
can be seen in the bottom left panel of \Fref{fig:cornerplot}, this is
indeed the case.

Differences in $\Delta\log{\aimf}$ reflect differences in merger
history.  Since mergers tend to decrease the effective IMF
normalization, we expect galaxies with the highest increase in stellar
mass to also show the largest changes in effective IMF. However this
is not observed to be the case, as can be seen from
\Fref{fig:cornerplot} (fourth row, third column).  If the amount of
accretion does not play a central role in setting the final IMF, then
the different change in IMF between low-mass and high-mass systems
must be set by differences in the type of mergers. To quantify this,
we measure the accreted stellar mass-weighted merger mass ratio,
$\left< \xi \right>_*$, defined as the average over all mergers
between $z=2$ and $z=0$ of the mass ratio between the accreted
satellites and the central halo weighted by the accreted stellar mass
with each merger:
\begin{equation}\label{eq:xieff}
\left< \xi \right>_* = \frac{\int dz \int d\xi \xi f(\xi, z)}{\int dz \int d\xi f(\xi, z)},
\end{equation}
with
\begin{equation}
f(\xi, z) \equiv \frac{\mathrm{d}^2N_{\mathrm{merg}}}{\mathrm{d}z \mathrm{d}\xi} \mathcal{R}_{*,h}(\xi \mhalo, z)\xi\mhalo.
\end{equation}
$\left< \xi \right>_*$ is plotted in \Fref{fig:cornerplot}. We see that $\left< \xi \right>_*$ increases for decreasing halo and stellar mass.
This means that for the most massive galaxies the bulk of the accreted stellar mass is brought in through mergers of relatively small mass ratio, while for less massive galaxies major mergers play a more important role.
Given our IMF prescription, where the heaviness of the IMF of satellites increases with mass, minor mergers produce a stronger decrease in the effective IMF of a galaxy than major mergers at fixed accreted stellar mass. 
This is seen in the bottom right panel of \Fref{fig:cornerplot}, where $\left< \xi \right>_*$ correlates with the change in effective IMF.
We then conclude that the shallowing of the $\msalp-\aimf$ or $\sigma-\aimf$ trends with time seen in our \Fref{fig:snap} is a result of the different relative importance of major and minor mergers in systems of different mass.

\subsection{Dependence of the IMF on global galaxy properties}

Another prediction of our model is the mixing between stellar mass and
velocity dispersion dependence of the IMF shown in \Fref{fig:tracks}.
The dependence of velocity dispersion, $a_\sigma$, seen at $z=0$ in
the ``$M_*$ model'' can be easily understood.  The ``$M_*$ model'' is
initialized with zero scatter around the $\msalp-\aimf$ relation at
$z=2$.  Since then, individual galaxies evolve roughly perpendicularly
to the $\msalp-\aimf$ relation (left hand panels of \Fref{fig:snap}),
so that by $z=1$ a significant scatter is introduced. At that point a
pure dependence of the IMF on stellar mass is no longer a good
description of the model, and a mixed dependence on both $M_*$ and
$\sigma$ is preferred.  In contrast in the ``$\sigma$ model'' galaxies
evolve parallel to the $\sigma-\aimf$ relation in place at $z=2$ and
very little scatter is added down to $z=0.3$, so that measuring the
scatter in IMF as a function of $z$ is potentially a powerful
diagnostic. No additional dependence on $M_*$ is required to describe
the data, hence the very small change in the parameters $a_*$ and
$a_\sigma$ for this model.  As discussed in \Sref{sect:results}
  however, this prediction relies on the assumption of parabolic
  orbits for the accreted satellites. If we allow for a milder
  decrease in velocity dispersion with dry mergers, we find a larger
  increase in the parameter $a_*$ and a correspondingly larger
  decrease in $a_\sigma$ for the ``$\sigma$ model''. Because of the
relatively rapid change in the coefficients $a_*$ and $a_\sigma$ with
time it is very difficult to determine from $z\sim0$ observations
which quantity between $\sigma$ and $M_*$ is the IMF fundamentally
dependent on. Current constraints from lensing and stellar dynamics
are unable to differentiate between the two models (see
\Fref{fig:tracks}).

An important assumption in our model is that all of the growth in
stellar mass between $z=2$ and $z=0$ is due to dry mergers. In other
terms, our model only applies to the population of quiescent galaxies
at $z=2$. This assumption complicates the comparison between
predictions from our model and observations, because the number
density of quiescent galaxies is observed to increase substantially
with time, in particular at $z>1$ \citep[e.g.][]{Ilb++13,
  Cas++13}. Depending on what the IMF of the newly quenched object is,
some results might change as a result of the so-called progenitor
bias. However: predictions on the $z<1$ evolution for the high-mass
end of the galaxy distribution should still hold, since the number
density of quiescent galaxies shows little evolution in that region of
the parameter space \citep{Lop++12}.

\section{Conclusions}\label{sect:concl} 

Using empirically motivated recipes for describing the IMF of massive
galaxies together with the dry merger evolution model developed by
\citet{Nip++12} we studied the time evolution of the effective IMF,
defined as the ratio between the true stellar mass and the stellar
mass one would infer assuming a Salpeter IMF, of a population of
massive galaxies from $z=2$ to $z=0$.  The models are set up to match
the observed correlations between IMF and stellar mass and velocity
dispersion in massive early-type galaxies at $z\sim0.3$.

Our model predicts a decrease in the effective IMF of individual objects with time as
galaxies merge with systems with a lighter IMF. This is seen in the
evolutionary tracks in \Fref{fig:snap}.
Trends between stellar mass
and velocity dispersion are qualitatively preserved by dry mergers,
but the slope of the correlation between stellar mass and effective
IMF becomes less steep with time.  

At fixed velocity
dispersion, the population of galaxies evolves with a constant or
increasing IMF normalization at increasing $z$.  This prediction
appears to be in slight tension with the few observational constraints
on the IMF available at $z>0.3$. However, there are systematic effects
that could be affecting observations the imporance of which has not
fully been assessed yet, most importantly the contribution of dark
matter and radial gradients in the IMF.  
Assuming that the tension is real and not the effect of systematics, the most plausible way to reconcile model and observations is to allow the effective IMF of satellite galaxies to be drawn from a different distribution, heavier at fixed galaxy properties, than that describing centrals.

Finally, we find that the relative
dependence of the IMF on stellar mass and velocity dispersion gets
mixed by dry mergers, making it difficult to observationally determine
the fundamental parameter(s) governing the IMF from $z\sim0$
measurements.

\section*{acknowledgments}

CN acknowledges financial support from PRIN MIUR 2010-2011,
project ‘The Chemical and Dynamical Evolution of the MilkyWay
and Local Group Galaxies’, prot. 2010LY5N2T.





\bibliographystyle{mnras}
\bibliography{references}

\begin{thebibliography}{}
\makeatletter
\relax
\def\mn@urlcharsother{\let\do\@makeother \do\$\do\&\do\#\do\^\do\_\do\%\do\~}
\def\mn@doi{\begingroup\mn@urlcharsother \@ifnextchar [ {\mn@doi@}
  {\mn@doi@[]}}
\def\mn@doi@[#1]#2{\def\@tempa{#1}\ifx\@tempa\@empty \href
  {http://dx.doi.org/#2} {doi:#2}\else \href {http://dx.doi.org/#2} {#1}\fi
  \endgroup}
\def\mn@eprint#1#2{\mn@eprint@#1:#2::\@nil}
\def\mn@eprint@arXiv#1{\href {http://arxiv.org/abs/#1} {{\tt arXiv:#1}}}
\def\mn@eprint@dblp#1{\href {http://dblp.uni-trier.de/rec/bibtex/#1.xml}
  {dblp:#1}}
\def\mn@eprint@#1:#2:#3:#4\@nil{\def\@tempa {#1}\def\@tempb {#2}\def\@tempc
  {#3}\ifx \@tempc \@empty \let \@tempc \@tempb \let \@tempb \@tempa \fi \ifx
  \@tempb \@empty \def\@tempb {arXiv}\fi \@ifundefined
  {mn@eprint@\@tempb}{\@tempb:\@tempc}{\expandafter \expandafter \csname
  mn@eprint@\@tempb\endcsname \expandafter{\@tempc}}}

\bibitem[\protect\citeauthoryear{{Auger}, {Treu}, {Gavazzi}, {Bolton},
  {Koopmans}  \& {Marshall}}{{Auger} et~al.}{2010a}]{Aug++10b}
{Auger} M.~W.,  {Treu} T.,  {Gavazzi} R.,  {Bolton} A.~S.,  {Koopmans}
  L.~V.~E.,   {Marshall} P.~J.,  2010a, \mn@doi [\apjl]
  {10.1088/2041-8205/721/2/L163}, \href
  {http://adsabs.harvard.edu/abs/2010ApJ...721L.163A} {721, L163}

\bibitem[\protect\citeauthoryear{{Auger}, {Treu}, {Bolton}, {Gavazzi},
  {Koopmans}, {Marshall}, {Moustakas}  \& {Burles}}{{Auger}
  et~al.}{2010b}]{Aug++10}
{Auger} M.~W.,  {Treu} T.,  {Bolton} A.~S.,  {Gavazzi} R.,  {Koopmans}
  L.~V.~E.,  {Marshall} P.~J.,  {Moustakas} L.~A.,   {Burles} S.,  2010b,
  \mn@doi [\apj] {10.1088/0004-637X/724/1/511}, \href
  {http://cdsads.u-strasbg.fr/abs/2010ApJ...724..511A} {724, 511}

\bibitem[\protect\citeauthoryear{{Barnab{\`e}}, {Czoske}, {Koopmans}, {Treu}
  \& {Bolton}}{{Barnab{\`e}} et~al.}{2011}]{Bar++11}
{Barnab{\`e}} M.,  {Czoske} O.,  {Koopmans} L.~V.~E.,  {Treu} T.,   {Bolton}
  A.~S.,  2011, \mn@doi [\mnras] {10.1111/j.1365-2966.2011.18842.x}, \href
  {http://adsabs.harvard.edu/abs/2011MNRAS.415.2215B} {415, 2215}

\bibitem[\protect\citeauthoryear{{Bastian}, {Covey}  \& {Meyer}}{{Bastian}
  et~al.}{2010}]{BCM10}
{Bastian} N.,  {Covey} K.~R.,   {Meyer} M.~R.,  2010, \mn@doi [\araa]
  {10.1146/annurev-astro-082708-101642}, \href
  {http://adsabs.harvard.edu/abs/2010ARA%26A..48..339B} {48, 339}

\bibitem[\protect\citeauthoryear{{Bekki}}{{Bekki}}{2013}]{Bek13}
{Bekki} K.,  2013, \mn@doi [\mnras] {10.1093/mnras/stt1735}, \href
  {http://adsabs.harvard.edu/abs/2013MNRAS.436.2254B} {436, 2254}

\bibitem[\protect\citeauthoryear{{Bezanson}, {van Dokkum}, {van de Sande},
  {Franx}  \& {Kriek}}{{Bezanson} et~al.}{2013}]{Bez++13}
{Bezanson} R.,  {van Dokkum} P.,  {van de Sande} J.,  {Franx} M.,   {Kriek} M.,
   2013, \mn@doi [\apjl] {10.1088/2041-8205/764/1/L8}, \href
  {http://adsabs.harvard.edu/abs/2013ApJ...764L...8B} {764, L8}

\bibitem[\protect\citeauthoryear{{Cappellari} et~al.,}{{Cappellari}
  et~al.}{2009}]{Cap++09}
{Cappellari} M.,  et~al., 2009, \mn@doi [\apjl] {10.1088/0004-637X/704/1/L34},
  \href {http://adsabs.harvard.edu/abs/2009ApJ...704L..34C} {704, L34}

\bibitem[\protect\citeauthoryear{{Cappellari} et~al.,}{{Cappellari}
  et~al.}{2012}]{Cap++12}
{Cappellari} M.,  et~al., 2012, \mn@doi [\nat] {10.1038/nature10972}, \href
  {http://adsabs.harvard.edu/abs/2012Natur.484..485C} {484, 485}

\bibitem[\protect\citeauthoryear{{Cassata} et~al.,}{{Cassata}
  et~al.}{2013}]{Cas++13}
{Cassata} P.,  et~al., 2013, \mn@doi [\apj] {10.1088/0004-637X/775/2/106},
  \href {http://adsabs.harvard.edu/abs/2013ApJ...775..106C} {775, 106}

\bibitem[\protect\citeauthoryear{{Chattopadhyay}, {De}, {Warlu}  \&
  {Chattopadhyay}}{{Chattopadhyay} et~al.}{2015}]{Cha++15}
{Chattopadhyay} T.,  {De} T.,  {Warlu} B.,   {Chattopadhyay} A.~K.,  2015,
  \mn@doi [\apj] {10.1088/0004-637X/808/1/24}, \href
  {http://adsabs.harvard.edu/abs/2015ApJ...808...24C} {808, 24}

\bibitem[\protect\citeauthoryear{{Conroy} \& {van Dokkum}}{{Conroy} \& {van
  Dokkum}}{2012}]{CvD12}
{Conroy} C.,  {van Dokkum} P.~G.,  2012, \mn@doi [\apj]
  {10.1088/0004-637X/760/1/71}, \href
  {http://adsabs.harvard.edu/abs/2012ApJ...760...71C} {760, 71}

\bibitem[\protect\citeauthoryear{{Dutton}, {Mendel}  \& {Simard}}{{Dutton}
  et~al.}{2012}]{Dut++12}
{Dutton} A.~A.,  {Mendel} J.~T.,   {Simard} L.,  2012, \mn@doi [\mnras]
  {10.1111/j.1745-3933.2012.01230.x}, \href
  {http://adsabs.harvard.edu/abs/2012MNRAS.422L..33D} {422, 33}

\bibitem[\protect\citeauthoryear{{Fakhouri}, {Ma}  \&
  {Boylan-Kolchin}}{{Fakhouri} et~al.}{2010}]{FMB10}
{Fakhouri} O.,  {Ma} C.-P.,   {Boylan-Kolchin} M.,  2010, \mn@doi [\mnras]
  {10.1111/j.1365-2966.2010.16859.x}, \href
  {http://adsabs.harvard.edu/abs/2010MNRAS.406.2267F} {406, 2267}

\bibitem[\protect\citeauthoryear{{Fontanot}, {De Lucia}, {Hirschmann},
  {Bruzual}, {Charlot}  \& {Zibetti}}{{Fontanot} et~al.}{2016}]{Fon++16}
{Fontanot} F.,  {De Lucia} G.,  {Hirschmann} M.,  {Bruzual} G.,  {Charlot} S.,
   {Zibetti} S.,  2016, preprint, \href
  {http://adsabs.harvard.edu/abs/2016arXiv160601908F} {} (\mn@eprint {arXiv}
  {1606.01908})

\bibitem[\protect\citeauthoryear{{Gargiulo} et~al.,}{{Gargiulo}
  et~al.}{2015a}]{GargiuloI++15}
{Gargiulo} I.~D.,  et~al., 2015a, \mn@doi [\mnras] {10.1093/mnras/stu2272},
  \href {http://adsabs.harvard.edu/abs/2015MNRAS.446.3820G} {446, 3820}

\bibitem[\protect\citeauthoryear{{Gargiulo}, {Saracco}, {Longhetti},
  {Tamburri}, {Lonoce}  \& {Ciocca}}{{Gargiulo} et~al.}{2015b}]{Gar++15}
{Gargiulo} A.,  {Saracco} P.,  {Longhetti} M.,  {Tamburri} S.,  {Lonoce} I.,
  {Ciocca} F.,  2015b, \mn@doi [\aap] {10.1051/0004-6361/201424235}, \href
  {http://adsabs.harvard.edu/abs/2015A%26A...573A.110G} {573, A110}

\bibitem[\protect\citeauthoryear{{Gavazzi}, {Treu}, {Rhodes}, {Koopmans},
  {Bolton}, {Burles}, {Massey}  \& {Moustakas}}{{Gavazzi}
  et~al.}{2007}]{Gav++07}
{Gavazzi} R.,  {Treu} T.,  {Rhodes} J.~D.,  {Koopmans} L.~V.~E.,  {Bolton}
  A.~S.,  {Burles} S.,  {Massey} R.~J.,   {Moustakas} L.~A.,  2007, \mn@doi
  [\apj] {10.1086/519237}, \href
  {http://adsabs.harvard.edu/abs/2007ApJ...667..176G} {667, 176}

\bibitem[\protect\citeauthoryear{{Guszejnov}, {Krumholz}  \&
  {Hopkins}}{{Guszejnov} et~al.}{2016}]{GKH16}
{Guszejnov} D.,  {Krumholz} M.~R.,   {Hopkins} P.~F.,  2016, \mn@doi [\mnras]
  {10.1093/mnras/stw315}, \href
  {http://adsabs.harvard.edu/abs/2016MNRAS.458..673G} {458, 673}

\bibitem[\protect\citeauthoryear{{Hennebelle} \& {Chabrier}}{{Hennebelle} \&
  {Chabrier}}{2011}]{H+C11}
{Hennebelle} P.,  {Chabrier} G.,  2011, \mn@doi [\apjl]
  {10.1088/2041-8205/743/2/L29}, \href
  {http://adsabs.harvard.edu/abs/2011ApJ...743L..29H} {743, L29}

\bibitem[\protect\citeauthoryear{{Hopkins}}{{Hopkins}}{2012}]{Hop12}
{Hopkins} P.~F.,  2012, \mn@doi [\mnras] {10.1111/j.1365-2966.2012.20731.x},
  \href {http://adsabs.harvard.edu/abs/2012MNRAS.423.2037H} {423, 2037}

\bibitem[\protect\citeauthoryear{{Hopkins}, {Bundy}, {Hernquist}, {Wuyts}  \&
  {Cox}}{{Hopkins} et~al.}{2010}]{Hop++10}
{Hopkins} P.~F.,  {Bundy} K.,  {Hernquist} L.,  {Wuyts} S.,   {Cox} T.~J.,
  2010, \mn@doi [\mnras] {10.1111/j.1365-2966.2009.15699.x}, \href
  {http://adsabs.harvard.edu/abs/2010MNRAS.401.1099H} {401, 1099}

\bibitem[\protect\citeauthoryear{{Ilbert} et~al.,}{{Ilbert}
  et~al.}{2013}]{Ilb++13}
{Ilbert} O.,  et~al., 2013, \mn@doi [\aap] {10.1051/0004-6361/201321100}, \href
  {http://adsabs.harvard.edu/abs/2013A%26A...556A..55I} {556, A55}

\bibitem[\protect\citeauthoryear{{Krumholz}}{{Krumholz}}{2011}]{Kru11}
{Krumholz} M.~R.,  2011, \mn@doi [\apj] {10.1088/0004-637X/743/2/110}, \href
  {http://adsabs.harvard.edu/abs/2011ApJ...743..110K} {743, 110}

\bibitem[\protect\citeauthoryear{{La Barbera}, {Ferreras}, {Vazdekis}, {de la
  Rosa}, {de Carvalho}, {Trevisan}, {Falc{\'o}n-Barroso}  \&
  {Ricciardelli}}{{La Barbera} et~al.}{2013}]{LaB++13}
{La Barbera} F.,  {Ferreras} I.,  {Vazdekis} A.,  {de la Rosa} I.~G.,  {de
  Carvalho} R.~R.,  {Trevisan} M.,  {Falc{\'o}n-Barroso} J.,   {Ricciardelli}
  E.,  2013, \mn@doi [\mnras] {10.1093/mnras/stt943}, \href
  {http://adsabs.harvard.edu/abs/2013MNRAS.433.3017L} {433, 3017}

\bibitem[\protect\citeauthoryear{{La Barbera}, {Vazdekis}, {Ferreras},
  {Pasquali}, {Cappellari}, {Mart{\'{\i}}n-Navarro}, {Sch{\"o}nebeck}  \&
  {Falc{\'o}n-Barroso}}{{La Barbera} et~al.}{2016}]{LaB++16}
{La Barbera} F.,  {Vazdekis} A.,  {Ferreras} I.,  {Pasquali} A.,  {Cappellari}
  M.,  {Mart{\'{\i}}n-Navarro} I.,  {Sch{\"o}nebeck} F.,   {Falc{\'o}n-Barroso}
  J.,  2016, \mn@doi [\mnras] {10.1093/mnras/stv2996}, \href
  {http://adsabs.harvard.edu/abs/2016MNRAS.457.1468L} {457, 1468}

\bibitem[\protect\citeauthoryear{{Leauthaud} et~al.,}{{Leauthaud}
  et~al.}{2012}]{Lea++12}
{Leauthaud} A.,  et~al., 2012, \mn@doi [\apj] {10.1088/0004-637X/744/2/159},
  \href {http://adsabs.harvard.edu/abs/2012ApJ...744..159L} {744, 159}

\bibitem[\protect\citeauthoryear{{L{\'o}pez-Sanjuan}
  et~al.,}{{L{\'o}pez-Sanjuan} et~al.}{2012}]{Lop++12}
{L{\'o}pez-Sanjuan} C.,  et~al., 2012, \mn@doi [\aap]
  {10.1051/0004-6361/201219085}, \href
  {http://adsabs.harvard.edu/abs/2012A%26A...548A...7L} {548, A7}

\bibitem[\protect\citeauthoryear{{Mart{\'{\i}}n-Navarro}, {La Barbera},
  {Vazdekis}, {Falc{\'o}n-Barroso}  \& {Ferreras}}{{Mart{\'{\i}}n-Navarro}
  et~al.}{2015}]{Mar++15}
{Mart{\'{\i}}n-Navarro} I.,  {La Barbera} F.,  {Vazdekis} A.,
  {Falc{\'o}n-Barroso} J.,   {Ferreras} I.,  2015, \mn@doi [\mnras]
  {10.1093/mnras/stu2480}, \href
  {http://adsabs.harvard.edu/abs/2015MNRAS.447.1033M} {447, 1033}

\bibitem[\protect\citeauthoryear{{Mason} et~al.,}{{Mason}
  et~al.}{2015}]{Mas++15}
{Mason} C.~A.,  et~al., 2015, \mn@doi [\apj] {10.1088/0004-637X/805/1/79},
  \href {http://adsabs.harvard.edu/abs/2015ApJ...805...79M} {805, 79}

\bibitem[\protect\citeauthoryear{{McConnell}, {Lu}  \& {Mann}}{{McConnell}
  et~al.}{2016}]{MLM16}
{McConnell} N.~J.,  {Lu} J.~R.,   {Mann} A.~W.,  2016, \mn@doi [\apj]
  {10.3847/0004-637X/821/1/39}, \href
  {http://adsabs.harvard.edu/abs/2016ApJ...821...39M} {821, 39}

\bibitem[\protect\citeauthoryear{{Naab}, {Johansson}  \& {Ostriker}}{{Naab}
  et~al.}{2009}]{NJO09}
{Naab} T.,  {Johansson} P.~H.,   {Ostriker} J.~P.,  2009, \mn@doi [\apjl]
  {10.1088/0004-637X/699/2/L178}, \href
  {http://adsabs.harvard.edu/abs/2009ApJ...699L.178N} {699, L178}

\bibitem[\protect\citeauthoryear{{Nagashima}, {Lacey}, {Okamoto}, {Baugh},
  {Frenk}  \& {Cole}}{{Nagashima} et~al.}{2005}]{Nag++05}
{Nagashima} M.,  {Lacey} C.~G.,  {Okamoto} T.,  {Baugh} C.~M.,  {Frenk} C.~S.,
   {Cole} S.,  2005, \mn@doi [\mnras] {10.1111/j.1745-3933.2005.00078.x}, \href
  {http://adsabs.harvard.edu/abs/2005MNRAS.363L..31N} {363, L31}

\bibitem[\protect\citeauthoryear{{Navarro}, {Frenk}  \& {White}}{{Navarro}
  et~al.}{1997}]{NFW97}
{Navarro} J.~F.,  {Frenk} C.~S.,   {White} S.~D.~M.,  1997, \apj, \href
  {http://adsabs.harvard.edu/abs/1997ApJ...490..493N} {490, 493}

\bibitem[\protect\citeauthoryear{{Newman}, {Ellis}, {Treu}  \&
  {Bundy}}{{Newman} et~al.}{2010}]{New++10}
{Newman} A.~B.,  {Ellis} R.~S.,  {Treu} T.,   {Bundy} K.,  2010, \mn@doi
  [\apjl] {10.1088/2041-8205/717/2/L103}, \href
  {http://adsabs.harvard.edu/abs/2010ApJ...717L.103N} {717, L103}

\bibitem[\protect\citeauthoryear{{Newman}, {Treu}, {Ellis}  \& {Sand}}{{Newman}
  et~al.}{2013}]{New++13}
{Newman} A.~B.,  {Treu} T.,  {Ellis} R.~S.,   {Sand} D.~J.,  2013, \mn@doi
  [\apj] {10.1088/0004-637X/765/1/25}, \href
  {http://adsabs.harvard.edu/abs/2013ApJ...765...25N} {765, 25}

\bibitem[\protect\citeauthoryear{{Newman}, {Ellis}  \& {Treu}}{{Newman}
  et~al.}{2015}]{New++15}
{Newman} A.~B.,  {Ellis} R.~S.,   {Treu} T.,  2015, \mn@doi [\apj]
  {10.1088/0004-637X/814/1/26}, \href
  {http://adsabs.harvard.edu/abs/2015ApJ...814...26N} {814, 26}

\bibitem[\protect\citeauthoryear{{Nipoti}, {Treu}, {Leauthaud}, {Bundy},
  {Newman}  \& {Auger}}{{Nipoti} et~al.}{2012}]{Nip++12}
{Nipoti} C.,  {Treu} T.,  {Leauthaud} A.,  {Bundy} K.,  {Newman} A.~B.,
  {Auger} M.~W.,  2012, \mn@doi [\mnras] {10.1111/j.1365-2966.2012.20749.x},
  \href {http://adsabs.harvard.edu/abs/2012MNRAS.422.1714N} {422, 1714}

\bibitem[\protect\citeauthoryear{{Offner}}{{Offner}}{2015}]{Off15}
{Offner} S.~S.~R.,  2015, preprint, \href
  {http://adsabs.harvard.edu/abs/2015arXiv151006027O} {} (\mn@eprint {arXiv}
  {1510.06027})

\bibitem[\protect\citeauthoryear{{Onodera} et~al.,}{{Onodera}
  et~al.}{2012}]{Ono++12}
{Onodera} M.,  et~al., 2012, \mn@doi [\apj] {10.1088/0004-637X/755/1/26}, \href
  {http://adsabs.harvard.edu/abs/2012ApJ...755...26O} {755, 26}

\bibitem[\protect\citeauthoryear{{Posacki}, {Cappellari}, {Treu}, {Pellegrini}
  \& {Ciotti}}{{Posacki} et~al.}{2015}]{Pos++15}
{Posacki} S.,  {Cappellari} M.,  {Treu} T.,  {Pellegrini} S.,   {Ciotti} L.,
  2015, \mn@doi [\mnras] {10.1093/mnras/stu2098}, \href
  {http://adsabs.harvard.edu/abs/2015MNRAS.446..493P} {446, 493}

\bibitem[\protect\citeauthoryear{{Posti}, {Nipoti}, {Stiavelli}  \&
  {Ciotti}}{{Posti} et~al.}{2014}]{Pos++14}
{Posti} L.,  {Nipoti} C.,  {Stiavelli} M.,   {Ciotti} L.,  2014, \mn@doi
  [\mnras] {10.1093/mnras/stu301}, \href
  {http://adsabs.harvard.edu/abs/2014MNRAS.440..610P} {440, 610}

\bibitem[\protect\citeauthoryear{{Salpeter}}{{Salpeter}}{1955}]{Sal55}
{Salpeter} E.~E.,  1955, \mn@doi [\apj] {10.1086/145971}, \href
  {http://adsabs.harvard.edu/abs/1955ApJ...121..161S} {121, 161}

\bibitem[\protect\citeauthoryear{{Schneider}, {Hogg}, {Marshall}, {Dawson},
  {Meyers}, {Bard}  \& {Lang}}{{Schneider} et~al.}{2015}]{Sch++15}
{Schneider} M.~D.,  {Hogg} D.~W.,  {Marshall} P.~J.,  {Dawson} W.~A.,  {Meyers}
  J.,  {Bard} D.~J.,   {Lang} D.,  2015, \mn@doi [\apj]
  {10.1088/0004-637X/807/1/87}, \href
  {http://adsabs.harvard.edu/abs/2015ApJ...807...87S} {807, 87}

\bibitem[\protect\citeauthoryear{{Sersic}}{{Sersic}}{1968}]{Ser68}
{Sersic} J.~L.,  1968, {Atlas de galaxias australes}.
Cordoba, Argentina: Observatorio Astronomico

\bibitem[\protect\citeauthoryear{{Shetty} \& {Cappellari}}{{Shetty} \&
  {Cappellari}}{2014}]{S+C14}
{Shetty} S.,  {Cappellari} M.,  2014, \mn@doi [\apjl]
  {10.1088/2041-8205/786/2/L10}, \href
  {http://adsabs.harvard.edu/abs/2014ApJ...786L..10S} {786, L10}

\bibitem[\protect\citeauthoryear{{Smith} \& {Lucey}}{{Smith} \&
  {Lucey}}{2013}]{S+L13}
{Smith} R.~J.,  {Lucey} J.~R.,  2013, \mn@doi [\mnras] {10.1093/mnras/stt1141},
  \href {http://adsabs.harvard.edu/abs/2013MNRAS.434.1964S} {434, 1964}

\bibitem[\protect\citeauthoryear{{Smith}, {Lucey}  \& {Conroy}}{{Smith}
  et~al.}{2015}]{SLC15}
{Smith} R.~J.,  {Lucey} J.~R.,   {Conroy} C.,  2015, \mn@doi [\mnras]
  {10.1093/mnras/stv518}, \href
  {http://adsabs.harvard.edu/abs/2015MNRAS.449.3441S} {449, 3441}

\bibitem[\protect\citeauthoryear{{Sonnenfeld}, {Nipoti}  \&
  {Treu}}{{Sonnenfeld} et~al.}{2014}]{SNT14}
{Sonnenfeld} A.,  {Nipoti} C.,   {Treu} T.,  2014, \mn@doi [\apj]
  {10.1088/0004-637X/786/2/89}, \href
  {http://adsabs.harvard.edu/abs/2014ApJ...786...89S} {786, 89}

\bibitem[\protect\citeauthoryear{{Sonnenfeld}, {Treu}, {Marshall}, {Suyu},
  {Gavazzi}, {Auger}  \& {Nipoti}}{{Sonnenfeld} et~al.}{2015}]{Son++15}
{Sonnenfeld} A.,  {Treu} T.,  {Marshall} P.~J.,  {Suyu} S.~H.,  {Gavazzi} R.,
  {Auger} M.~W.,   {Nipoti} C.,  2015, \mn@doi [\apj]
  {10.1088/0004-637X/800/2/94}, \href
  {http://adsabs.harvard.edu/abs/2015ApJ...800...94S} {800, 94}

\bibitem[\protect\citeauthoryear{{Spiniello}, {Trager}, {Koopmans}  \&
  {Conroy}}{{Spiniello} et~al.}{2014}]{Spi++14}
{Spiniello} C.,  {Trager} S.,  {Koopmans} L.~V.~E.,   {Conroy} C.,  2014,
  \mn@doi [\mnras] {10.1093/mnras/stt2282}, \href
  {http://adsabs.harvard.edu/abs/2014MNRAS.438.1483S} {438, 1483}

\bibitem[\protect\citeauthoryear{{Spiniello}, {Barnab{\`e}}, {Koopmans}  \&
  {Trager}}{{Spiniello} et~al.}{2015}]{Spi++15}
{Spiniello} C.,  {Barnab{\`e}} M.,  {Koopmans} L.~V.~E.,   {Trager} S.~C.,
  2015, \mn@doi [\mnras] {10.1093/mnrasl/slv079}, \href
  {http://adsabs.harvard.edu/abs/2015MNRAS.452L..21S} {452, L21}

\bibitem[\protect\citeauthoryear{{Tinker}, {Kravtsov}, {Klypin}, {Abazajian},
  {Warren}, {Yepes}, {Gottl{\"o}ber}  \& {Holz}}{{Tinker}
  et~al.}{2008}]{Tin++08}
{Tinker} J.,  {Kravtsov} A.~V.,  {Klypin} A.,  {Abazajian} K.,  {Warren} M.,
  {Yepes} G.,  {Gottl{\"o}ber} S.,   {Holz} D.~E.,  2008, \mn@doi [\apj]
  {10.1086/591439}, \href {http://adsabs.harvard.edu/abs/2008ApJ...688..709T}
  {688, 709}

\bibitem[\protect\citeauthoryear{{Tortora}, {Romanowsky}  \&
  {Napolitano}}{{Tortora} et~al.}{2013}]{TRN13}
{Tortora} C.,  {Romanowsky} A.~J.,   {Napolitano} N.~R.,  2013, \mn@doi [\apj]
  {10.1088/0004-637X/765/1/8}, \href
  {http://adsabs.harvard.edu/abs/2013ApJ...765....8T} {765, 8}

\bibitem[\protect\citeauthoryear{{Tortora}, {Napolitano}, {Saglia},
  {Romanowsky}, {Covone}  \& {Capaccioli}}{{Tortora} et~al.}{2014}]{Tor++14}
{Tortora} C.,  {Napolitano} N.~R.,  {Saglia} R.~P.,  {Romanowsky} A.~J.,
  {Covone} G.,   {Capaccioli} M.,  2014, \mn@doi [\mnras]
  {10.1093/mnras/stu1712}, \href
  {http://adsabs.harvard.edu/abs/2014MNRAS.445..162T} {445, 162}

\bibitem[\protect\citeauthoryear{{Treu}, {Auger}, {Koopmans}, {Gavazzi},
  {Marshall}  \& {Bolton}}{{Treu} et~al.}{2010}]{Tre++10}
{Treu} T.,  {Auger} M.~W.,  {Koopmans} L.~V.~E.,  {Gavazzi} R.,  {Marshall}
  P.~J.,   {Bolton} A.~S.,  2010, \mn@doi [\apj]
  {10.1088/0004-637X/709/2/1195}, \href
  {http://adsabs.harvard.edu/abs/2010ApJ...709.1195T} {709, 1195}

\bibitem[\protect\citeauthoryear{{de Vaucouleurs}}{{de
  Vaucouleurs}}{1948}]{deV48}
{de Vaucouleurs} G.,  1948, Annales d'Astrophysique, \href
  {http://adsabs.harvard.edu/abs/1948AnAp...11..247D} {11, 247}

\bibitem[\protect\citeauthoryear{{van Dokkum} et~al.,}{{van Dokkum}
  et~al.}{2010}]{vDo++10}
{van Dokkum} P.~G.,  et~al., 2010, \mn@doi [\apj]
  {10.1088/0004-637X/709/2/1018}, \href
  {http://adsabs.harvard.edu/abs/2010ApJ...709.1018V} {709, 1018}

\bibitem[\protect\citeauthoryear{{van de Sande} et~al.,}{{van de Sande}
  et~al.}{2013}]{vdS++13}
{van de Sande} J.,  et~al., 2013, \mn@doi [\apj] {10.1088/0004-637X/771/2/85},
  \href {http://adsabs.harvard.edu/abs/2013ApJ...771...85V} {771, 85}

\bibitem[\protect\citeauthoryear{{van der Wel}, {Holden}, {Zirm}, {Franx},
  {Rettura}, {Illingworth}  \& {Ford}}{{van der Wel} et~al.}{2008}]{vdW++08}
{van der Wel} A.,  {Holden} B.~P.,  {Zirm} A.~W.,  {Franx} M.,  {Rettura} A.,
  {Illingworth} G.~D.,   {Ford} H.~C.,  2008, \mn@doi [\apj] {10.1086/592267},
  \href {http://adsabs.harvard.edu/abs/2008ApJ...688...48V} {688, 48}

\makeatother
\end{thebibliography}

\appendix
\onecolumn
\section{Hierarchical Bayesian inference}\label{sect:appendix}
We use a similar hierarchical Bayesian inference method as the one
used by S15 to fit for the distribution in IMF of the same sample of
galaxies. The method can be summarized as follows.  We assume that the
density profile of early-type galaxies can be described by the sum of
a stellar component with a de Vaucouleurs profile and a dark-matter
component with an NFW \citep{NFW97} profile.  We
parametrize this model in terms of the Salpeter stellar mass $\msalp$,
effective radius $\reff$, projected dark-matter mass within a cylinder
of radius $5$ kpc $\mdm$, dark-matter scale radius $r_s$ and effective
IMF $\aimf$.  Following S15 we assume that the effective radius is
known exactly and we fix $r_s = 10\reff$ for simplicity. The exact
value of $r_s$ has little impact on lensing and dynamics measurements,
more sensistive to the mass distribution at the scale of a few kpc,
much smaller than typical values of $r_s$ predicted by numerical
simulations.  The free parameters of each object are then $\msalp$,
$\mdm$ and $\aimf$. For a more compact notation we label this triplet $\indpar$.
\begin{equation}
\indpar \equiv(\msalp, \mdm, \aimf).
\end{equation}
From these parameters we calculate the model
central velocity dispersion, $\sigmaee$, using the spherical Jeans
equation and assuming isotropic stellar velocity distribution.
Though a simplification, the spherical isotropic model has been shown
to match results obtained with a more accurate dynamical modeling
\citep{Bar++11}.

We assume that the parameters of each galaxy are drawn from a
distribution for the population of galaxies, described by a set of
{\em hyper-parameters} $\hyperp$:
\begin{equation}
\indpari \sim \pr(\indpari | \hyperp),
\end{equation}
where the subscript $i$ indicates the $i$-th galaxy in the sample.
We model this distribution as the product of four terms:
\begin{equation}\label{eq:hyperp}
\pr(\indpari | \hyperp) = \mathcal{G}_*(\msalpi|\hyperp) \mathcal{G}_\sigma(\sigmaeei|\msalp, \hyperp) \mathcal{G}_{\mathrm{DM}}(\mdmi|\msalp,\sigmaeei,\hyperp) \mathcal{G}_{\mathrm{IMF}}(\aimfi|\msalp,\sigmaeei,\hyperp),
\end{equation}
where each term is a log-normal distribution respectively in $\msalp$, $\sigmaee$ (which is a well-defined function of $\indpar$), $\mdm$ and $\aimf$.
More precisely,
\begin{equation}\label{eq:mstarterm}
\mathcal{G}_*(\msalpi|\hyperp) = \frac{1}{\sqrt{2\pi}\sigma_*}\exp{\left\{-\frac{(\log{\msalpi} - \mu_*)^2}{2\sigma_*^2}\right\}}.
\end{equation}
\begin{equation}
\mathcal{G}_*(\sigmaeei|\hyperp) = \frac{1}{\sqrt{2\pi}\sigma_\sigma}\exp{\left\{-\frac{(\log{\sigmaeei} - \mu_\sigma(z_i, \msalpi))^2}{2\sigma_\sigma^2}\right\}},
\end{equation}
where
\begin{equation}
\mu_\sigma(\msalpi) = \mu_{\sigma,0} + \zeta_\sigma(z_i - 0.3) + \beta_\sigma(\log{\msalpi} - 11.5).
\end{equation}
\begin{equation}
\mathcal{G}_*(\mdmi|\hyperp) = \frac{1}{\sqrt{2\pi}\sigma_{\mathrm{DM}}}\exp{\left\{-\frac{(\log{\mdmi} - \mu_{\mathrm{DM}}(z_i, \msalpi, \sigmaeei))^2}{2\sigma_{\mathrm{DM}}^2}\right\}},
\end{equation}
where
\begin{equation}
\mu_{\mathrm{DM}}(\msalpi, \sigmaeei) = \mu_{\mathrm{DM},0} + \zeta_{\mathrm{DM}}(z_i - 0.3) + \beta_{\mathrm{DM}}(\log{\msalpi} - 11.5) + \xi_{\mathrm{DM}}(\log{\sigmaeei} - 2.4).
\end{equation}
\begin{equation}
\mathcal{G}_*(\aimfi|\hyperp) = \frac{1}{\sqrt{2\pi}\sigma_{\mathrm{IMF}}}\exp{\left\{-\frac{(\log{\aimfi} - \mu_{\mathrm{IMF}}(z, \msalpi, \sigmaeei))^2}{2\sigma_{\mathrm{IMF}}^2}\right\}},
\end{equation}
where
\begin{equation}
\mu_{\mathrm{IMF}}(\msalpi, \sigmaeei) = a_z(z_i - 0.3) + a_*(\log{\msalpi} - 11.5) + a_\sigma(\log{\sigmaeei} - 2.4) + \log{\alpha_{\mathrm{IMF},0}}.
\end{equation}
Since the lenses in the S15 study are collected from two different
surveys, SLACS and SL2S, we allow for different values in the mean and
dispersion of the stellar mass distribution in \Eref{eq:mstarterm}: $\mu_*^{\mathrm{SLACS}}$, $\mu_*^{\mathrm{SL2S}}$, $\sigma_*^{\mathrm{SLACS}}$ and $\sigma_*^{\mathrm{SL2S}}$.

In summary the distribution of the parameters describing individual galaxies, \Eref{eq:hyperp}, is described by 18 hyper-parameters:
\begin{equation}
\hyperp = \left\{\mu_*^{\mathrm{SLACS}}, \mu_*^{\mathrm{SL2S}}, \sigma_*^{\mathrm{SLACS}}, \sigma_*^{\mathrm{SL2S}}, \mu_{\sigma,0}, \zeta_\sigma, \beta_\sigma, \sigma_\sigma, \mu_{\mathrm{DM}, 0}, \zeta_{\mathrm{DM}}, \beta_{\mathrm{DM}}, \xi_{\mathrm{DM}}, \sigma_{\mathrm{DM}}, \log{\alpha_{\mathrm{IMF}, 0}}, a_z, a_*, a_\sigma, \sigma_{\mathrm{IMF}} \right\}.
\end{equation}
We wish to infer the posterior probability distribution of these hyper-parameters given the data $\datad$.
The data consists of a measurement of the Einstein radius, velocity dispersion and stellar mass from stellar population synthesis analysis for each object.
From Bayes theorem, 
\begin{equation}
\pr(\hyperp|\datad) \propto \pr(\hyperp)\pr(\datad|\hyperp),
\end{equation}
where $\pr(\hyperp)$ is the prior on the hyper-parameters and $\pr(\datad|\hyperp)$ the likelihood of observing the data given the values of the hyper-parameters.
This latter term is the product over all galaxies
\begin{equation}
\pr(\datad|\hyperp) = \prod_i \pr(\datadi|\hyperp).
\end{equation}
Each term in this product can in turn be expanded as
\begin{equation}\label{eq:integrals}
\pr(\datadi|\hyperp) = \int d\indpari \pr(\datadi|\indpari) \pr(\indpari|\hyperp).
\end{equation}
We assume flat priors on all hyper-parameters.
We then sample the posterior probability distribution with a Markov Chain Monte Carlo. 
The integrals in \Eref{eq:integrals} are evaulated via importance sampling and Monte Carlo integration following \citet{Sch++15}.
The inferred values of the hyper-parameters with $1-\sigma$ uncertainties are reported in \Tref{tab:hyperp}.
\begin{table*}
 \caption{Full model, corresponding to \Fref{fig:sl2scorner}. Median and $68\%$ limits on the posterior probability function of each hyper-parameter, marginalized over the other parameters.}
 \label{tab:hyperp}
 \begin{tabular}{lcl}
 \hline
 & & Parameter description \\
 \hline
 $\mu_*^{\mathrm{SLACS}}$ & $11.58 \pm 0.03$ & Mean $\log{\msalp}$, SLACS sample\\
$\sigma_*^{\mathrm{SLACS}}$ & $0.19 \pm 0.02$ & Scatter in $\log{\msalp}$, SLACS sample\\
$\mu_*^{\mathrm{SL2S}}$ & $11.50 \pm 0.05$ & Mean $\log{\msalp}$, SL2S sample\\
$\sigma_*^{\mathrm{SL2S}}$ & $0.23 \pm 0.04$ & Scatter in $\log{\msalp}$, SL2S sample\\
$\mu_{\sigma,0}$ & $2.39 \pm 0.01$ & Mean $\log{\sigma}$ at $z=0.3$ and $\log{\msalp}=11.5$\\
$\zeta_\sigma$ & $-0.01 \pm 0.03$ & Linear dependence of $\log{\sigma}$ on $z$\\
$\beta_\sigma$ & $0.19 \pm 0.03$ & Linear dependence of $\log{\sigma}$ on $\log{\msalp}$\\
$\sigma_\sigma$ & $0.04 \pm 0.01$ & Scatter in $\log{\sigma}$\\
$\mu_{\mathrm{DM},0}$ & $10.60 \pm 0.09$ & Mean $\log{\mdm}$ at $z=0.3$, $\log{\msalp}=11.5$ and $\log{\sigma}=2.4$\\
$\zeta_{\mathrm{DM}}$ & $1.13 \pm 0.27$ & Linear dependence of $\log{\mdm}$ on $z$\\
$\beta_{\mathrm{DM}}$ & $0.13 \pm 0.25$ & Linear dependence of $\log{\mdm}$ on $\log{\msalp}$\\
$\xi_{\mathrm{DM}}$ & $-0.77 \pm 0.72$ & Linear dependence of $\log{\mdm}$ on $\log{\sigma}$\\
$\sigma_{\mathrm{DM}}$ & $0.25 \pm 0.05$ & Scatter in $\log{\mdm}$\\
$\log{\alpha_{\mathrm{IMF},0}}$ & $0.06 \pm 0.02$ & Mean $\log{\aimf}$ at $z=0.3$, $\log{\msalp}=11.5$ and $\log{\sigma}=2.4$\\
$a_z$ & $-0.20 \pm 0.09$ & Linear dependence of $\log{\aimf}$ on $z$\\
$a_*$ & $0.10 \pm 0.10$ & Linear dependence of $\log{\aimf}$ on $\log{\msalp}$\\
$a_\sigma$ & $0.49 \pm 0.39$ & Linear dependence of $\log{\aimf}$ on $\log{\sigma}$\\
$\sigma_{\mathrm{IMF}}$ & $0.03 \pm 0.01$ & Scatter in $\log{\aimf}$\\

 \hline
 \end{tabular}
\end{table*}

As described in \Sref{sect:obs}, we then repeated the analysis by fixing $a_*=0$. 
In this second analysis we also fix $\beta_{\mathrm{DM}}=0$.
The inferred values of the hyper-parameters with $1-\sigma$ uncertainties are reported in \Tref{tab:sigmaonlyhp}.
\begin{table*}
 \caption{Model with $a_*=0$, corresponding to \Fref{fig:sigmaonlycorner}.  Median and $68\%$ limits on the posterior probability function of each hyper-parameter, marginalized over the other parameters.}
 \label{tab:sigmaonlyhp}
 \begin{tabular}{lcl}
 \hline
 & & Parameter description \\
 \hline
 $\mu_*^{\mathrm{SLACS}}$ & $11.59 \pm 0.03$ & Mean $\log{\msalp}$, SLACS sample\\
$\sigma_*^{\mathrm{SLACS}}$ & $0.19 \pm 0.02$ & Scatter in $\log{\msalp}$, SLACS sample\\
$\mu_*^{\mathrm{SL2S}}$ & $11.50 \pm 0.05$ & Mean $\log{\msalp}$, SL2S sample\\
$\sigma_*^{\mathrm{SL2S}}$ & $0.24 \pm 0.04$ & Scatter in $\log{\msalp}$, SL2S sample\\
$\mu_{\sigma,0}$ & $2.39 \pm 0.01$ & Mean $\log{\sigma}$ at $z=0.3$ and $\log{\msalp}=11.5$\\
$\zeta_\sigma$ & $-0.01 \pm 0.04$ & Linear dependence of $\log{\sigma}$ on $z$\\
$\beta_\sigma$ & $0.18 \pm 0.02$ & Linear dependence of $\log{\sigma}$ on $\log{\msalp}$\\
$\sigma_\sigma$ & $0.05 \pm 0.00$ & Scatter in $\log{\sigma}$\\
$\mu_{\mathrm{DM},0}$ & $10.63 \pm 0.07$ & Mean $\log{\mdm}$ at $z=0.3$, $\log{\msalp}=11.5$ and $\log{\sigma}=2.4$\\
$\zeta_{\mathrm{DM}}$ & $1.12 \pm 0.24$ & Linear dependence of $\log{\mdm}$ on $z$\\
$\xi_{\mathrm{DM}}$ & $-0.52 \pm 0.68$ & Linear dependence of $\log{\mdm}$ on $\log{\sigma}$\\
$\sigma_{\mathrm{DM}}$ & $0.25 \pm 0.05$ & Scatter in $\log{\mdm}$\\
$\log{\alpha_{\mathrm{IMF},0}}$ & $0.06 \pm 0.01$ & Mean $\log{\aimf}$ at $z=0.3$, $\log{\msalp}=11.5$ and $\log{\sigma}=2.4$\\
$a_z$ & $-0.19 \pm 0.09$ & Linear dependence of $\log{\aimf}$ on $z$\\
$a_\sigma$ & $0.81 \pm 0.21$ & Linear dependence of $\log{\aimf}$ on $\log{\sigma}$\\
$\sigma_{\mathrm{IMF}}$ & $0.04 \pm 0.01$ & Scatter in $\log{\aimf}$\\

 \hline
 \end{tabular}
\end{table*}

\end{document}